
%
\input amstex
\input amsppt.sty

\CenteredTagsOnSplits
\NoBlackBoxes
\def\today{\ifcase\month\or
 January\or February\or March\or April\or May\or June\or
 July\or August\or September\or October\or November\or December\fi
 \space\number\day, \number\year}
\define\({\left(}
\define\){\right)}

\define\Aut{\operatorname{Aut}}
\define\CC{{\Bbb C}}

\define\Hom{\operatorname{Hom}}
\define\Map{\operatorname{Map}}

\define\RR{{\Bbb R}}

\define\Tr{\operatorname{Tr}}
\define\ZZ{{\Bbb Z}}
\define\[{\left[}
\define\]{\right]}

\define\chiup{\raise.5ex\hbox{$\chi$}}
\define\cir{S^1}

\define\exertag #1#2{\removelastskip\bigskip\medskip\eightpoint\noindent%
\hbox{\rm\ignorespaces#2\unskip} #1.\ }

\define\inv{^{-1}}
\define\mstrut{^{\vphantom{1*\prime y}}}
\define\protag#1 #2{#2\ #1}

\define\res#1{\negmedspace\bigm|_{#1}}
\define\temsquare{\raise3.5pt\hbox{\boxed{ }}}

\define\theprotag#1 #2{#2~#1}

\define\zmod#1{\ZZ/#1\ZZ}

\define\As{A^*}
\define\Eas{\Es_\lambda}
\define\Es{E^*}
\define\HG#1{\Hom(\pi_1 #1,\Gamma)}
\define\Inv{\operatorname{Inv}}

\define\Obj{\operatorname{Obj}}
\define\Pb{[P]}
\define\Pcuttil{\widetilde{P\cut}}
\define\Qbcut{[Q\cut]}
\define\Qb{[Q]}
\define\RZ{\RR/\ZZ}
\define\Rb{{\bold R}}
\define\TT{{\Bbb T}}

\define\ac#1#2{S_X(#1,#2)}
\define\ah{\hat{\alpha}}
\define\as{a^*}
\define\bX{\partial X}
\define\bal{\boldsymbol\lambda }
\define\bbcyli{[\bcyl]\inv}
\define\bbcyln#1{[\bcyl_#1]}
\define\bbcylp{[\bcyl']}
\define\bbcyl{[\bcyl]}
\define\bcyl{T}
\define\bfld#1{{\Cal C}^\prime_{#1}}
\define\bfldb#1{\overline{\bfld{#1}}}
\define\bfldbu#1{\overline{\vphantom{\Cal{L}\mstrut _Y}\Cal{C}^\prime_{#1}}}
\define\bone{\bold {1}}
\define\bs{b^*}
\define\cs{c^*}
\define\cut{^{\text{cut}}}
\define\eac#1#2{e^{2\pi iS_{#1}({#2})}}
\define\eint#1#2{\exp\( 2\pi i\int_{#1}{#2}\)}
\define\esp{e^{*\prime}}
\define\es{e^*}
\define\fld#1{{\Cal C}_{#1}}
\define\fldb#1{\overline{\fld{#1}}}
\define\gh{\hat{g}}
\define\id{\operatorname{id}}
\define\iline#1{I_{#1,\alpha}}
\define\intline#1#2{I_{#1,#2}}
\define\op{^{\text{op}}}
\define\phibar{\bar{\varphi}}

\define\ptt{\operatorname{pt}}

\define\supp{\operatorname{supp}}
\define\te#1#2{\ell _{\langle #1,#2 \rangle}}
\define\tes#1#2{\te #1#2^*}
\define\thetatil{\tilde{\theta}}
\define\tpi{2\pi i}
\define\voll{\operatorname{vol}}

\define\zo{[0,1]}

\NoRunningHeads
\loadbold
\refstyle{A}
\widestnumber\key{DVVV}
 \pretitle{$$\boxed{\boxed{\text{REVISED VERSION}}}$$\par\vskip 1.5 pc}
 \topmatter
 \title\nofrills Chern-Simons Theory with Finite Gauge Group \endtitle
 \author Daniel S. Freed\\ Frank Quinn \endauthor
 \thanks The first author is supported by NSF grant DMS-8805684, an Alfred P.
Sloan Research Fellowship, a Presidential Young Investigators award, and by
the O'Donnell Foundation.  The second author is supported by NSF grant
DMS-9207973.\endthanks
 \affil Department of Mathematics, University of Texas at Austin\\
Department of Mathematics, Virginia Polytechnical Institute \endaffil
 \address Department of Mathematics, University of Texas, Austin, TX
78712\endaddress
 \email dafr\@math.utexas.edu \endemail
 \address Department of Mathematics, VPI \& SU, Blacksburg, VA 24061\endaddress
 \email quinn\@vtmath.math.vt.edu \endemail
 \date December 11, 1992\enddate
\endtopmatter

\document

A typical course in quantum field theory begins with a thorough examination
of a ``toy model'', usually the $\phi^4$ theory. Our purpose here is to
provide a detailed description of a ``toy model'' for {\it topological\/}
quantum field theory, suitable for use as a foundation for more sophisticated
developments. We carry through all the steps of the path integral
quantization: start with a lagrangian, construct the classical action,
construct a measure, and do the integral. When the gauge group is finite the
``path integral'' reduces to a finite sum. This remark makes it clear that
the analytical difficulties simplify enormously, and that there should be no
essential problem in carrying out the process. Many interesting features
remain, however. The algebraic and topological structure are essentially
unchanged, and are much clearer when not overshadowed by the analysis. And
even the analysis does not entirely disappear: the details of the
construction of the state spaces requires a much more precise formulation of
the classical theory than is usually given, and reveals some incompleteness
in the understanding of the classical theory for continuous Lie
groups~\cite{F1}.

Chern-Simons theory with finite gauge group was introduced by Dijkgraaf and
Witten \cite{DW}, who essentially cataloged the possible lagrangians and gave
some sample calculations. Special cases were considered by Segal \cite{S2}
and Kontsevich \cite{K}.  More abstract and mathematically-oriented versions
are considered in \cite{Q1}, \cite{Q2}, and connections with the
representation-theoretic approach of Reshetikhin and Turaev are described by
Ferguson~\cite{Fg} and Yetter~\cite{Y}.

Let $\Gamma$ denote the gauge group, which we assume to be finite.  In~\S1
and~\S2 we carry out the quantization: describe the lagrangians, classical
actions, and path integrals. The fields in this version are regular covering
spaces $P\to X$ whose group of deck transformations is $\Gamma$. The action
is a (torsion) characteristic number associated to a class in
$H^3(B\Gamma;\RR/\ZZ)$. We represent this class in singular cohomology by a
singular cocycle, and use this write the action (1.6) as an integral over
$X$. (The cochain plays the role of the lagrangian.) We caution that the
resulting theory depends in a subtle way on this choice of cocycle: a
different choice gives a theory which is isomorphic in an appropriate sense,
but the isomorphism between the two is not canonical: it depends on further
choices. We must also make careful sense of integration of singular cocycles
over manifolds with boundary; this is explained in Appendix B.  The classical
theory is somewhat unusual in that the action on manifolds with boundary is
not a number, but rather an element in a complex line determined by the
restriction of the field to the boundary.  These lines, which properly belong
to the hamiltonian theory, are the source of much of the structure in the
theory.  We defer some of the details of the classical theory to~\cite{F1}
and~\cite{F3}, which explores the classical Chern-Simons theory for {\it
arbitrary\/} compact gauge groups.  The quantization in~\S{2} is
straightforward.  Of general interest is~\theprotag{2.4} {Lemma}, which
explains the behavior of symmetry groups and measures under gluing,
and~\thetag{2.18}, which proves the gluing law in the quantum theory.

One distinguishing feature of the Chern-Simons theory (with any gauge group)
in $2+1$~dimensions is its relationship to conformal field theory in
$1+1$~dimensions.  As a result one constructs quantum Hilbert spaces not only
for closed surfaces, but also for surfaces with boundary,\footnote{This
requires fixing certain boundary data at the outset.  The role of these
choices is investigated in~\cite{F5}.} and these vector spaces obey a gluing
law related to the gluing law of the path integral.  We explore these ideas,
which we term ``modular structure'', in~\S{3}.  (The name derives from the
term ``modular functor'', which was coined by Segal~\cite{S1}.)  The
quantization of the cylinder produces a semisimple coalgebra~$A$, and the
quantum spaces attached to surfaces with boundary are $A$-comodules.  The
gluing law~\thetag{3.26} is expressed in terms of a cotensor product.  (We
review the relevant algebra in Appendix~A.)  In rational conformal field
theories (see~\cite{MS}, for example) one has a set of ``labels'' which index
the conformal blocks.  They appear in the definition of a modular
functor~\cite{S1} and also in most treatments of Chern-Simons
theory~\cite{W}.  Here they index the irreducible corepresentations of~$A$,
as we discuss in~\S{4}.  The gluing law for surfaces with labeled boundary
is~\thetag{4.11}.  The numerical factors which describe the behavior of the
inner products also appear in Kevin Walker's careful treatment~\cite{Wa} of
the $SU(2)$ ~theory.  The Verlinde algebra~\cite{V} is easily derived from
this gluing law.

We remark that all of these theories have ``level'' zero.\footnote{As a
result there are no central extensions of diffeomorphism groups, as there are
in theories with continuous gauge group.} The cohomology class in
$H^3(B\Gamma ;\RZ)\cong H^4(B\Gamma ;\ZZ)$, which in the $SU(2)$~theory gives
the level, is pure torsion.  In the simplest theory (the ``untwisted
theory'') this class vanishes.  Then the path integral essentially counts
representations of the fundamental group into~$\Gamma $.  We include some
computations for the untwisted theory in~\S{5}.  Here we also find an {\it
algebra\/} structure on~$A$, and so $A$~becomes a Hopf algebra.  It is the
dual of the Hopf algebra considered by Dijkgraaf, Pasquier, and
Roche~\cite{DPR}, and plays the role of the quantum group in these finite
theories.  In~\S{5} we also calculate the action of~$SL(2;\ZZ)$ on the
quantum space attached to the torus in arbitrary twisted theories.

In a subsequent paper~\cite{F5} we investigate more fully the relationship
between the path integral and the Hopf algebras, or quantum groups, which
arise in the field theory.  There we show how to interpret the path integral
over manifolds of lower dimension, and how the quantum group emerges from
those considerations.

The axioms for topological quantum field theory were first formulated by
Atiyah~\cite{A}, who also gives a mathematical perspective on the subject.
These ideas are developed much further in~\cite{Q1}, ~\cite{Q2} where more
examples are pursued.  There is an expository account of gluing laws in
topological field theory in~\cite{F4}.

We warmly thank Joseph Bernstein, Robert Dijkgraaf, Gary Hamrick, Graeme
Segal, Karen Uhlenbeck, and Ed Witten for many conversations touching on the
subject matter of this paper.  We understand that Joseph Bernstein and David
Kazhdan independently did work similar to that discussed here.

\newpage
\head
\S{1} Classical Theory
\endhead
\comment
lasteqno 1@ 19
\endcomment

The basic data of a Lagrangian classical field theory is a space of
fields~$\fld X$ for each spacetime~$X$ and a functional $S_X\:\fld X\to\RR$
called the {\it action\/}.  The fields and action must be {\it local\/},
i.e., computable by cutting and pasting.  Usually the action is the integral
of a locally defined differential form over spacetime, and so is local by
standard properties of the integral.  Also, the action usually depends on a
metric on spacetime.  Here the theory is topological so no metric is needed.
The formal properties of a classical topological field theory are carefully
stated in \theprotag{1.7} {Theorem}.  (See ~\cite{S1}, ~\cite{A}, and
\theprotag{2.13} {Theorem} for analogous axiomatizations of topological
quantum field theory.)  In the theory we consider here a field is a finite
regular covering space whose group of deck transformations is a fixed finite
group~$\Gamma $.  Unlike fields in ordinary theories, which are usually
functions, these fields (covering spaces) have automorphisms.  This
complication leads to our use of categorical language, which is designed to
handle mathematical objects with automorphisms.  We must keep track of
automorphisms to make proper sense of cutting and pasting.  The action for
spacetimes without boundary is a characteristic class which takes its values
in~$\RZ$.  For spacetimes with boundary the characteristic class is not
standardly defined.  We introduce an integration theory for singular cocycles
in Appendix~B, and express the characteristic class as an integral.  Hence as
in standard field theories our action is an integral over spacetime.  We
remark that we can also define the action using a refinement of integration
in de Rham theory~\cite{F2}.  In $2+1$~dimensions the classical field theory
in this section is a special case of the classical Chern-Simons theory which
is defined for arbitrary compact Lie groups~\cite{F1}, ~\cite{F3}.

Fix a finite group~$\Gamma $.  A {\it classifying space\/}~$B\Gamma$
for~$\Gamma$ is a connected space with homotopy groups $\pi
_1(B\Gamma,*)\cong \Gamma$ and $\pi _j(B\Gamma,*)=0,\, j\ge 2$, for any
basepoint~$*\in B\Gamma$.  In the topology literature this is also called an
Eilenberg-MacLane space~$K(\Gamma,1)$.  We construct a classifying space as
follows.  Denote $\Gamma =\{g_1,\dots ,g_N\}$.  Suppose $H$~is an infinite
dimensional separable complex Hilbert space, and let $E\Gamma $~be the
Stiefel manifold of ordered orthonormal sets $\{v_{g_1},\dots
,v_{g_N}\}\subset H$ labeled by the elements of~$\Gamma $.  Then $g\in \Gamma
$~acts on $\{v_{g_1},\dots ,v_{g_N}\}$ by permutation:
  $$ \{v_{g_1},\dots ,v_{g_N}\}\cdot g = \{v_{g_1g},\dots ,v_{g_Ng}\}.
     $$
This defines a free right action of~$\Gamma $ on~$E\Gamma $, and we set
$B\Gamma =E\Gamma /\Gamma $.  Since $E\Gamma $~is contractible, the
quotient~$B\Gamma $ has the requisite properties.

Now a {\it principal $\Gamma $~bundle\/} over a manifold~$M$ is a
manifold~$P$ with a free right $\Gamma $ action such that $P/\Gamma =M$.
Notice that $P\to M$ is a finite regular covering space.  A map of $\Gamma
$~bundles $\varphi \:P'\to P$ is a smooth map which commutes with the $\Gamma
$~action.  There is an induced map $\phibar\:M'\to M$ on the quotients.  If
$M'=M$ and $\phibar=\id$, then we term~$\varphi $ a {\it morphism\/}.  Notice
that any morphism has an inverse.  Hence if there exists a morphism $\varphi
\:P'\to P$, we say that $P'$~is equivalent to~$P$.  A morphism $\varphi
\:P\to P$ is an {\it automorphism\/} (or {\it gauge transformation\/} or {\it
deck transformation\/}).  Let $\fld M$~be the category of principal $\Gamma
$~bundles over~$M$ and bundle morphisms.  Since every morphism is invertible,
this category is a groupoid.  Finally, denote by~$\fldb M$ the set of
equivalence classes of $\Gamma $~bundles over~$M$.  It is a finite set if
$M$~is compact.  In fact, if $M$~is connected then there is a natural
identification
  $$ \fldb M \cong \Hom\bigl( \pi _1(M,m),\Gamma \bigr)\bigm/ \Gamma
     \tag{1.1} $$
for any basepoint~$m\in M$, where $\Gamma $~acts by conjugation.  The
identification is via monodromy.

If $P\to M$ is a $\Gamma $~bundle, then there exists a bundle map $F\:P\to
E\Gamma $.  Such an ~$F$ is called a {\it classifying map\/} for~$P$.  (This
explains the term `classifying space.')  To construct~$F$ note that the
$\Gamma $~bundle $(P\times E\Gamma )/\Gamma \to M$ has contractible fibers.
Hence there exist sections $M\to (P\times E\Gamma )/\Gamma $, or equivalently
$\Gamma $~maps $P\to E\Gamma $.  All sections are homotopic, whence all
classifying maps are homotopic through $\Gamma $~maps.

After these preliminaries we are ready to define the classical theory.  Fix
an integer~$d\ge 0$.  The `spacetimes' in our theory have dimension~$d+1$ and
the `spaces' have dimension~$d$.\footnote{One can consider CW complexes (of
arbitrary dimension) instead of manifolds of a fixed dimension~\cite{Q1},
{}~\cite{Q2}, ~\cite{K}.  We restrict to manifolds to make contact with
standard physical theories.} The ingredients of such a theory are a space of
fields~$\fld X$ attached to every compact oriented spacetime~$X$ and an
action functional~$S_X$ defined on~$\fld X$.  In our theory we take ~$\fld X$
to be the discrete space of principal $\Gamma $~bundles defined above.

Now the action.  Fix a class\footnote{Notice that $H^{d+1}(B\Gamma ;\RZ)\cong
H^{d+2}(B\Gamma )$ since the real cohomology of~$B\Gamma $ is trivial.  We
could consider complex cohomology classes in~$H^{d+1}(B\Gamma ;\CC/\ZZ)$, and
then the resulting theories are {\it unitary\/} only if the class is
real.}~$\alpha \in H^{d+1}(B\Gamma ;\RZ)$ and a singular cocycle $\ah\in
C^{d+1}(B\Gamma ;\RZ)$ which represents~$\alpha $.  The simplest theory has
$\alpha =\ah=0$, in which case the action (and so the whole classical theory)
is completely trivial.  Suppose $X$~is a {\it closed\/} oriented
$(d+1)$-manifold.  Fix~$P\in \fld X$ and let $F \:P\to E\Gamma $ be a
classifying map.  There is a quotient map $\overline{F}\:X\to B\Gamma $.  Set
  $$ S_X(P) = \alpha \bigl( \overline{F}_*[X]\bigr)\in \RZ \tag{1.2}$$
where $[X]\in H_{d+1}(X)$ is the fundamental class.  Since all classifying
maps for~$P$ are homotopic through $\Gamma $~maps, the action~\thetag{1.2}
does not depend on the choice of~$F $.  Also, the action on closed manifolds
only depends on the cohomology class~$\alpha $, not on the particular
cocycle~$\ah$.  We often write the action as~$\eac XP\in \TT$; it takes
values in the circle group~$\TT$ of complex numbers with unit norm.

The action on manifolds with boundary is not a number, but rather is an
element in a complex line associated to the boundary.  We first
abstract the construction of vector spaces in situations where one must make
choices.  We like to call this the {\it invariant section
construction\/}.\footnote{This is a special case of a general construction
(the ``limit'' or ``inverse limit'' or ``projective limit'') in category
theory~\cite{Mac}.} Suppose that the set of possible choices and isomorphisms
of these choices forms a groupoid~$\Cal{C}$.  Let $\Cal{L}$~be the category
whose objects are metrized complex lines (one dimensional inner product
spaces) and whose morphisms are unitary isomorphisms.  Suppose we have a
functor $\Cal{F}\:\Cal{C}\to\Cal{L}$.  Define $V_{\Cal{F}}$ to be the inner
product space of {\it invariant\/} sections of the functor~$\Cal{F}$: An
element $v\in V_{\Cal{F}}$ is a collection $\{v(C)\in \Cal{F}(C)\}_{C\in
\Obj(\Cal{C})}$ such that if $C_1@>\psi >>C_2$ is a morphism, then
$\Cal{F}(\psi )v(C_1) = v(C_2)$.  Suppose $\Cal{C}$~is {\it connected\/},
that is, there is a morphism between any two objects.  Then $\dim
V_{\Cal{F}}=0$ or $\dim V_{\Cal{F}}=1$, the latter occurring if and only if
$\Cal{F}$~{\it has no holonomy\/}, i.e., $\Cal{F}(\psi )=\id$ for every
automorphism $C@>\psi >> C$.  We will apply this construction many times in
this paper, both in the classical theory and in its quantization.

Return to the finite group~$\Gamma $ and cocycle $\ah\in C^{d+1}(B\Gamma
;\RZ)$.  Let $Q\to Y$ be a principal $\Gamma $~bundle over a closed oriented
$d$-manifold~$Y$.  We now define a metrized line~$L_Q$.  Consider the
category~$\Cal{C}_Q$ whose objects are classifying maps $f\:Q\to E\Gamma $
for~$Q$ and whose morphisms $f @>h>> f'$ are homotopy classes rel boundary of
$\Gamma $-homotopies $h\:\zo\times Q\to E\Gamma $ from~$f$ to~$f'$.  The
category~$\Cal{C}_Q$ is connected since any two classifying maps are $\Gamma
$-homotopic.  Define a functor $\Cal{F}_Q\:\Cal{C}_Q\to \Cal{L}$ as follows.
Let
  $$ \Cal{F}_Q(f)= I_{Y,\bar{f}^*\ah} \tag{1.3}$$
be the metrized integration line of \theprotag{B.1} {Proposition}, and
$\Cal{F}_Q(f @>h>> f')$ the map
  $$ \exp\( 2\pi i\int_{\zo\times Y}{\bar{h}^*\ah}\) \:
     I_{Y,\bar{f}^*\ah} \longrightarrow I_{Y,\bar{f}'{}^*\ah}. \tag{1.4}$$
The integral in~\thetag{1.4} is also defined in \theprotag{B.1}
{Proposition}.  That $\Cal{F}_Q$~is a functor follows from \thetag{B.2}.
That \thetag{1.4}~only depends on the homotopy class of~$h$ follows
from~\thetag{B.3}.  Furthermore, $\Cal{F}_Q$~has no holonomy.\footnote{Here
it is crucial that our classifying maps are defined on the bundles and not
just on the base spaces.} For if $f @>h>> f$ is an automorphism, then
$h$~determines a bundle map $h\: S^1\times Q\to E\Gamma $ by gluing.  Since
the bundle $ S^1\times Q\to S^1\times Y$ extends over~$ D^2\times Y$, so too
does the map $\bar{h}\: S^1\times Y\to B\Gamma $ extend to a map
$\overline{H}\:D^2\times Y\to B\Gamma $, and so
  $$ \eint{S^1\times Y}{\bar{h}^*\ah} = \eint{D^2\times Y}{\overline{H}^*\ah}
     = 1   \tag{1.5}$$
by \theprotag{B.1(e)} {Proposition}, which proves that $\Cal{F}_Q$~has no
holonomy.  Let $L_Q$~be the metrized line of invariant sections
of~$\Cal{F}_Q$.

Now suppose $P\to X$ is a $\Gamma $~bundle over a compact oriented
$(d+1)$-manifold~$X$.  For each classifying map $F\:P\to E\Gamma $ consider
the quantity
  $$ e^{2\pi iS_X(P,F)} = \eint X{\overline{F}^*\ah}\in
     L_{\bX,\overline{\partial F}^*\ah}. \tag{1.6}$$
Here $\partial F=F\res{\partial P}$ is the restriction of~$F$ to the
boundary.  Suppose $F'$~is another classifying map for~$P$.  Choose a
homotopy $K\:\zo\times P\to E\Gamma $ with $K\res{\{0\}\times P}=F$ and
$K\res{\{1\}\times P}=F'$.  Let $k=K\res{\zo\times \partial P}$ be the
induced homotopy from~$\partial F$ to~$\partial F'$.  Denote $I=[0,1]$.
There is a product class $[I]\times [X]\in H_{d+2}\bigl(I\times X,\partial
(I\times X)\bigr)$ with
  $$ \partial ( [I]\times [X] ) = \partial [I]\times [X] \,\cup\, (-1)^{\dim I}
     [I]\times \partial [X] =\{1\}\times [X] \,\cup\, -\{0\}\times [X] \,\cup\,
     -[I]\times [\partial X]. $$
Choose representative chains~$i,x$ for~$[I]$ and~$[X]$.  Then since $\ah$~is
closed,
  $$ 0=\overline{F}'{}^*\ah(x) - \overline{F}^*\ah(x) -
     \bar{k}^*\ah(i\times \partial x). $$
Applying $\exp(2\pi i\cdot )$ we see by \thetag{1.4} that
  $$ \Cal{F}_{\partial P}\( \partial F @>k>> \partial F' \) e^{2\pi
     iS_X(P,F)} =  e^{2\pi iS_X(P,F')}. $$
Thus \thetag{1.6}~determines an invariant section of~$\Cal{F}_{\partial P}$,
i.e., an element (of unit norm)
  $$ \eac X P\in L_{\partial P}. $$
This defines the action on manifolds with boundary.

The following theorem expresses what we mean by the statement ``$S_X$~is the
action of a local Lagrangian field theory.''

     \proclaim{\protag{1.7} {Theorem}}
 Let $\Gamma $~be a finite group and $\ah\in C^{d+1}(B\Gamma ;\RZ)$ a
cocycle.  Then the assignments
  $$ \alignedat 2
     Q &\longmapsto L_Q ,&&\qquad Q \in \fld Y,\\
     P &\longmapsto \eac XP,&&\qquad P \in \fld X\endalignedat
     \tag{1.8}$$
defined above for closed oriented $d$-manifolds~$Y$ and compact oriented
$(d+1)$-manifolds~$X$ satisfy:\newline
 \rom(a\rom)\ \rom({\it Functoriality\/}\rom)\ If $\psi \:Q'\to Q$ is a bundle
map covering an orientation preserving diffeomorphism $\overline{\psi}\:Y'\to
Y$, then there is an induced isometry
  $$ \psi _*\:L_{Q'} \longrightarrow L_{Q} \tag{1.9}$$
and these compose properly.  If $\varphi \:P'\to P$ is a bundle map covering
an orientation preserving diffeomorphism $\phibar\:X'\to X$, then
  $$ (\partial \varphi )_*\(\eac {X'}{P'}\) = \eac {X}{P},
     \tag{1.10}$$
where $\partial \varphi \:\partial P'\to\partial P$ is the induced map over
the boundary.\newline
 \rom(b\rom)\ \rom({\it Orientation\/}\rom)\ There is a natural isometry
  $$ L_{Q ,-Y} \cong \overline{L_{Q ,Y}}, \tag{1.11}$$
and
  $$ \eac{-X}P = \overline{\eac XP}. \tag{1.12}$$
 \rom(c\rom)\ \rom({\it Additivity\/}\footnote{Although \thetag{1.14}~looks
like a multiplicative property, it expresses the additivity of the classical
action~$S_X$.  However, $S_X$~is not defined if $\partial X\not= \emptyset $,
which is why we use the exponential notation~$e^{2\pi iS_X(\cdot )}$.}\rom)
\ If $Y=Y_1\sqcup Y_2$ is a disjoint union, and $Q _i$~are bundles over~$Y_i$,
then there is a natural isometry
  $$ L_{Q _1\sqcup Q _2} \cong L_{Q _1}\otimes L_{Q _2}.
     \tag{1.13}$$
If $X=X_1\sqcup X_2$ is a disjoint union, and $P _i$~are bundles
over~$X_i$, then
  $$ \eac{X_1\sqcup X_2}{P _1\sqcup P _2} = \eac{X_1}{P _1}
     \otimes  \eac{X_2}{P _2}. \tag{1.14}$$
 \rom(d\rom)\ \rom({\it Gluing\/}\rom)\ Suppose $Y\hookrightarrow X$ is a
closed oriented codimension one submanifold and $X\cut$~is the manifold
obtained by cutting~$X$ along~$Y$.  Then $\partial X\cut = \partial X\sqcup Y
\sqcup -Y$.  Suppose $P $~is a bundle over~$X$, $P \cut$~the induced bundle
over~$X\cut$, and $Q $~the restriction of~$P $ to~$Y$.  Then
  $$ \eac XP = \Tr_Q \( \eac{X\cut}{P \cut}\), \tag{1.15}$$
where $\Tr_Q $ is the contraction
  $$ \Tr_Q \:L_{\partial P \cut} \cong  L_{\partial P }\otimes
     L_Q  \otimes \overline{L_Q}\longrightarrow L_{\partial P }
     \tag{1.16}$$
using the hermitian metric on~$L_Q $.
     \endproclaim

Several comments are in order.  We allow the empty set~$\emptyset $ as a
manifold: $L_\emptyset =\CC$ and~$S_\emptyset \equiv 0$.  From a functorial
point of view, (a)~implies that $Q \mapsto L_Q $ defines a functor
 $$\fld Y\to\Cal{L} $$
and that each~$X$ determines an invariant section~$\eac X\cdot $ of the
composite functor $\fld X\to\fld{\partial X}\to\Cal{L}$, where the first
arrow is restriction to the boundary.  The invariance of the
action~\thetag{1.10} on closed manifolds~$X$ means that if~$P'\cong P$, then
$S_X(P')=S_X(P)$.  Hence the action passes to a function
  $$ S_X\:\fldb X\longrightarrow \RZ. \tag{1.17}$$
Bundle morphisms over compact oriented $d$-manifolds~$Y$ act on the
corresponding lines, via~\thetag{1.9}.  So there is a line bundle
  $$ \Cal{L}_Y\to\fld Y \tag{1.18}$$
with a lift of the morphisms in~$\fld Y$.  (Since $\fld Y$~is a discrete set
of points, $\Cal{L}_Y$~is a discrete union of lines.)  If $X$~is a compact
oriented $(d+1)$-manifold, there is an induced line bundle $\Cal{L}_X\to\fld
X$, obtained by pulling back~$\Cal{L}_{\bX}$ via the restriction map $\fld
X\to\fld{\bX}$.  The action~$\eac X\cdot $ is an invariant section of
$\Cal{L}_X\to\fld X$.  In particular, the group of automorphisms~$\Aut P$
of~$P\to X$ acts on the line over~$P\in \fld X$.  If this action is
nontrivial, then $\eac XP=0$.  \theprotag{1.7} {Theorem} expresses in a
(necessarily) complicated way the fact that ``$S_X$''~is a local functional
of local fields defined as the integral of a local expression (c), ~(d); is
invariant under symmetries of the fields~(a); and changes sign under
orientation reversal~(b).

     \demo{Proof}
 The proof of \theprotag{1.7} {Theorem} is straightforward but tedious.  For
example, to construct the map~\thetag{1.9}, choose a classifying map
$f'\:Q'\to E\Gamma $ for~$Q'$, and let $f\circ \psi\inv\:Q\to E\Gamma $ be
the induced classifying map for~$Q$.  Fix a representative~$y'\in C_d(Y')$
of~$[Y']$, and let $y=\bar{\psi}_*(y')\in C_d(Y)$ be the corresponding
representative of~$[Y]$.  These choices determine
trivializations~$L_{Q'}\cong \CC$ and~$L_{Q}\cong \CC$.  Relative to these
trivializations we define~$\psi _*$ in~\thetag{1.9} to be the identity.  A
routine check shows that this is independent of the choices and composes
properly.  The constructions of the isometries~\thetag{1.11}
and~\thetag{1.13} are similar.  Equations~\thetag{1.10}, \thetag{1.12},
\thetag{1.14}, and~\thetag{1.16} follow from the corresponding properties of
the integral (\theprotag{B.1} {Proposition}).
      \enddemo

The metrized line bundle~\thetag{1.18} passes to a (possibly degenerate)
metrized line bundle
  $$ \overline{\Cal{L}_Y}\longrightarrow \fldb Y \tag{1.19}$$
over the finite set of equivalence classes.  The fiber~$L_{\Qb}$ is the space
of invariant sections of the functor~$Q\mapsto L_Q$ as $Q$~ranges over the
equivalence class~$\Qb$.  If $\Aut Q$~acts nontrivially on~$L_Q$ then
$L_{\Qb}=0$; if $\Aut Q$~acts trivially then $\dim L_{\Qb}=1$.

\newpage
\head
\S{2} Quantum Theory
\endhead
\comment
lasteqno 2@ 18
\endcomment

There is one crucial piece of data which must be added to a classical field
theory to define path integral quantization: a measure on the space of
fields.  In many quantum field theories this is only done formally.  In our
theory this is easy to do precisely since the space of fields is discrete.
We simply count each bundle according to the number of its
automorphisms~\thetag{2.1}.  The path integral for closed spacetimes is then
defined directly~\thetag{2.9} as the integral of the action over the space of
(equivalence classes of) fields, whereas for manifolds with boundary the path
integral is a function of the field on the boundary~\thetag{2.12}.  In usual
field theories the fields on space (and on spacetime) have continuous
parameters; then extra geometry\footnote{In geometric quantization this is
called a `polarization'.} is introduced to carry out `canonical
quantization', and again this is only a formal procedure in many cases of
interest.  Here no extra geometry is needed since the space of fields is
discrete.  The quantum Hilbert space is the space of all functions of
fields~\thetag{2.10} (which respect the symmetry group); here it is finite
dimensional.  The path integral and quantum Hilbert spaces satisfy a set of
axioms we spell out in \theprotag{2.13} {Theorem} (cf.~\cite{Se}, ~\cite{A}).
They mostly follow from the corresponding properties of the classical theory
(\theprotag{1.7} {Theorem}) and the gluing properties of the measure
(\theprotag{2.4} {Lemma}).  We emphasize that since the configuration
space~$\fldb X$ of equivalence classes of fields is finite, the path integral
reduces to a finite sum.

We introduce a measure~$\mu $ on the collection~$\fld M$ of principal $\Gamma
$~bundles over any manifold~$M$.  Namely, set
  $$ \mu _P = \frac{1}{\#\Aut P}, \tag{2.1}$$
where $\Aut P$ is the group of automorphisms of~$P$.  If~$P'\cong P$, then
$\Aut P'\cong \Aut P$, so that~$\mu _{P'}= \mu _P$.  Hence there is an
induced measure on the set of equivalence classes~$\fldb M$.

Suppose $M$~is a manifold with boundary, and $Q\in \fld{\partial M}$ is a
$\Gamma $~bundle over the boundary.  Define the category
  $$ \fld M(Q) = \{\langle P,\theta \rangle: P\in \fld
     M,\;\text{$\theta\:\partial P\to Q$ is an isomorphism} \} $$
of bundles over~$M$ whose boundary has a specified isomorphism to~$Q$.  A
morphism $\varphi \:\langle P',\theta ' \rangle\to\langle P,\theta \rangle$
is an isomorphism $\varphi \:P'\to P$ such that $\theta '=\theta \circ
\partial \varphi $.  Two elements $\langle P',\theta ' \rangle,\langle
P,\theta \rangle\in \fld M(Q)$ are equivalent if there exists a
morphism~$\langle P',\theta '\rangle\to \langle P,\theta \rangle$.  We denote
the set of equivalence classes by~$\fldb M(Q)$.  Equation~\thetag{2.1}
defines a measure on~$\fld M(Q)$, where we now interpret `$\Aut \langle
P,\theta \rangle$' in the sense just described.  Notice that such
automorphisms are trivial over components of~$M$ with nonempty boundary.
Again the measure passes to the quotient~$\fldb M(Q)$.  Finally, if $\psi
\:Q'\to Q$ is a morphism, there is an induced measure preserving map
  $$ \psi _*\:\fldb M(Q')\longrightarrow \fldb M(Q). \tag{2.2}$$

Next, we investigate the behavior of these measures under cutting and
pasting.  Suppose $N\hookrightarrow M$ is an oriented codimension one
submanifold and $M\cut$ the manifold obtained by cutting~$M$ along~$N$.  Then
$\partial M\cut = \partial M\sqcup N\sqcup -N$.  Fix a bundle $Q\to N$.  Then
$\fld{M\cut}(Q\sqcup Q)$~is the category of triples~$\langle P\cut,\theta
_1,\theta _2\rangle $, where $P\cut\to M\cut$ is a $\Gamma $~bundle and
$\theta _i\:P\cut\res N\to Q$ are isomorphisms over the two copies of~$N$
in~$M\cut$.  Consider the gluing map
  $$ \aligned
     g_Q\:\fldb{M\cut}(Q\sqcup Q) &\longrightarrow \fldb M,\\
     \langle P\cut,\theta _1,\theta _2\rangle &\longmapsto P\cut\bigm/
     (\theta _1=\theta _2)\endaligned \tag{2.3}$$
on equivalence classes.

     \proclaim{\protag{2.4} {Lemma}}
 The gluing map~$g_Q$ satisfies\newline
 \rom(a\rom)\  $g_Q$~maps onto the set of bundles over~$M$ whose restriction
to~$N$
is isomorphic to~$Q$.\newline
 \rom(b\rom)\ Let $\phi \in \Aut Q$ act on $\langle P\cut,\theta _1,\theta _2
\rangle\in \fld{M\cut}(Q\sqcup Q)$ by
  $$ \phi \cdot \langle  P\cut,\theta _1,\theta _2\rangle  = \langle
     P\cut,\phi \circ \theta _1, \phi \circ \theta _2\rangle . $$
Then the stabilizer of this action at~$\langle P\cut,\theta _1,\theta
_2\rangle $ is the image $\Aut P\to\Aut Q$ determined by the~$\theta _i$,
where $P = g_Q(\langle P\cut,\theta _1,\theta _2\rangle )$.  \newline
 \rom(c\rom)\ There is an induced action on equivalence classes
$\fldb{M\cut}(Q\sqcup
Q)$, and $\Aut Q$~acts transitively on~$g\inv _Q([P])$ for any~$[P]\in
\fldb{M}$.  \newline
 \rom(d\rom)\ For all $[P]\in \fldb M$ we have
  $$ \mu _{[P]} = \voll\bigl( g\inv _Q([P]) \bigr)\cdot \mu
     ^{\vphantom{-1}}_{Q}. \tag{2.5}$$
     \endproclaim

     \demo{Proof}
 If $P\to M$ is a bundle and $\theta \:P\res N\to Q$ an isomorphism, then
$g_Q\bigl( \langle P\cut,\theta ,\theta \rangle \bigr) \cong P$, where
$P\cut$~is the pullback of~$P$ under the gluing map $M\cut\to M$.  This
proves~(a).  If $g_Q\bigl( \langle P\cut,\theta_1,\theta_2\rangle \bigr)
\cong g_Q\bigl( \langle \Pcuttil,\thetatil_1,\thetatil_2\rangle \bigr)$, then
there exists an isomorphism $\varphi \:P\cut\to\Pcuttil$ such that $
\thetatil^{\vphantom{1}}_1\varphi \theta _1\inv =\thetatil^{\vphantom{1}}_2
\varphi \theta _2\inv \in \Aut Q$.  Call this element~$\phi $.  Then $\varphi
$~determines an isomorphism $\phi \cdot \langle P\cut,\theta _1,\theta
_2\rangle \cong \langle \Pcuttil,\thetatil_1,\thetatil_2\rangle $.  This
proves~(c).  If $\langle P\cut,\theta _1,\theta _2\rangle = \langle
\Pcuttil,\thetatil_1,\thetatil_2\rangle $, then $\varphi $~determines an
element of~$\Aut P$, and $\phi $~is the restriction of this element to~$Q$.
This proves~(b).  Now the exact sequence
  $$ 1 @>>> \Aut \langle P\cut,\theta _1,\theta_2\rangle @>>> \Aut P @>>>
     \Aut Q $$
together with~(b) and~(c) imply
  $$ \#\Aut P = \#\Aut \langle P\cut,\theta _1,\theta_2\rangle  \,\frac{\#\Aut
     Q}{\#\bigl( g_Q\inv (P)\bigr)}. $$
This is equivalent to~\thetag{2.5}.
     \enddemo

\flushpar
 We remark that this computation is valid both for automorphisms which are the
identity over~$\partial M$ and automorphisms which are unrestricted
on~$\partial M$.

Now we are ready to carry out the quantization.  The quantum theory assigns a
complex inner product space~$E(Y)$ to every closed oriented $d$-manifold~$Y$
and a vector~$Z_X\in E(\bX)$ to every compact oriented $(d+1)$-manifold~$X$.
If $X$~is closed, then $Z_X$~is a complex number.  We begin with the $\alpha
=\ah=0$ theory, which is surely the simplest quantum field theory that one
could imagine.  For $X$~closed the path integral is
  $$ Z_X = \int_{\fldb X}d\mu (\Pb) = \voll(\fldb X). \tag{2.6}$$
The vector space attached to~$Y$ is
  $$ E(Y) = L^2(\fldb Y). \tag{2.7}$$
If $X$~is a manifold with boundary, then the path integral is
  $$ Z_X(Q) = \int_{\fldb X(Q)}d\mu (\Pb) = \voll\bigl( \fldb X(Q)
     \bigr),\qquad Q\in \fld {\bX}. \tag{2.8}$$
Since \thetag{2.2}~is a measure preserving map, $Z_X(Q)$~only depends on the
equivalence class of~$Q$, so defines an element of~$E(\partial X)$ as
desired.

The formul\ae\ for the twisted case ($\ah\not= 0$) are obtained by
substituting the nontrivial action~$\eac X\cdot $ for the trivial action~$1$
in~\thetag{2.6}--\thetag{2.8}.  Thus for $X$~closed we define the partition
function
  $$ Z_X = \int_{\fldb X}d\mu (\Pb)\, \eac X{\Pb}, \tag{2.9}$$
where $S_X(\Pb)$~is the action~\thetag{1.17} on the quotient.  We emphasize
that the (path) integral in~\thetag{2.9} is a finite sum.
For $Y$~a closed oriented $d$-manifold we have the possibly degenerate
metrized line bundle~\thetag{1.19}.  Set
  $$ E(Y) = L^2 (\fldb Y,\overline{\Cal{L}_Y}). \tag{2.10}$$
In other words,  $E(Y)$~is the space of invariant sections of the functor
  $$ \aligned
     \Cal{F}_Y\:\fld Y &\longrightarrow \Cal{L},\\
     Q &\longmapsto L_Q.\endaligned $$
If $v,v'$~are invariant sections, then
  $$ (v,v') _{E(Y)} = \sum\limits_{\Qb\in \fldb Y} \mu _Q\,
     \bigl( v(Q), v'(Q) \bigr) _{L_Q}, \tag{2.11}$$
where $Q$~is a bundle in the equivalence class~$\Qb$.  Finally, if $X$~is a
manifold with boundary, set
  $$ Z_X(Q) = \int_{\fldb X(Q)} d\mu (\Pb)\,\eac X{\Pb} \in L_Q,\qquad Q\in
     \fld{\bX}. \tag{2.12}$$
A simple application of~\thetag{2.2} and~\thetag{1.10} proves that
\thetag{2.12}~defines an invariant section of the functor~ $\Cal{F}_{\bX}$,
so an element in~$E(\bX)$.

The following theorem expresses what we mean by the statement ``$E(Y)$~
and $Z_X$~ define a (unitary) topological quantum field theory.''  It is
essentially the list of axioms in~\cite{A}.

     \proclaim{\protag{2.13} {Theorem}}
 Let $\Gamma $~be a finite group and $\ah\in C^{d+1}(B\Gamma ;\RZ)$ a
cocycle.  Then the assignments
  $$ \aligned
     Y &\longmapsto E(Y) ,\\
     X &\longmapsto Z_X,\endaligned  \tag{2.14}$$
defined above for closed oriented $d$-manifolds~$Y$ and compact oriented
$(d+1)$-manifolds~$X$ satisfy:\newline
 \rom(a\rom)\ \rom({\it Functoriality\/}\rom)\ Suppose $f\:Y'\to Y$ is an
orientation preserving diffeomorphism.  Then there is an induced isometry
  $$ f_*\:E(Y') \longrightarrow E(Y)  \tag{2.15}$$
and these compose properly.  If $F\:X'\to X$ is an orientation preserving
diffeomorphism, then
  $$ (\partial F)_*(Z_{X'}) = Z_X,  \tag{2.16}$$
where $\partial F\:\partial X'\to\partial X$ is the induced map over the
boundary.\newline
 \rom(b\rom)\ \rom({\it Orientation\/}\rom)\ There is a natural isometry
  $$ E(-Y) \cong \overline{E(Y)},  $$
and
  $$ Z_{-X} = \overline{Z_X}.  $$
 \rom(c\rom)\ \rom({\it Multiplicativity\/}\rom)\ If $Y=Y_1\sqcup Y_2$ is a
disjoint union, then there is a natural isometry
  $$ E(Y _1\sqcup Y _2) \cong E(Y _1)\otimes E(Y _2).  $$
If $X=X_1\sqcup X_2$ is a disjoint union, then
  $$ Z_{X_1\sqcup X_2} = Z_{X_1} \otimes  Z_{X_2}.  $$
 \rom(d\rom)\ \rom({\it Gluing\/}\rom)\ Suppose $Y\hookrightarrow X$ is a
closed oriented codimension one submanifold and $X\cut$~is the manifold
obtained by cutting~$X$ along~$Y$.  Write $\partial X\cut = \partial X\sqcup
Y \sqcup -Y$.  Then
  $$ Z_ X = \Tr_Y ( Z_{X\cut}),  \tag{2.17}$$
where $\Tr_Y $ is the contraction
  $$ \Tr_Y \:E(\partial X \cut) \cong  E(\partial X )\otimes
     E(Y)  \otimes \overline{E(Y)}\longrightarrow E(\partial X )
     $$
using the hermitian metric on~$E(Y) $.
     \endproclaim

     \demo{Proof}
 The map~$f$ induces a measure preserving functor $f^*\:\fld Y\to \fld{Y'}$,
which lifts to $\tilde{f}^*\:L_Y\to L_{Y'}$ in view of~\thetag{1.9}.  Then
\thetag{2.15}~is the induced pullback on invariant sections.  The
diffeomorphism invariance~\thetag{2.16} follows from~\thetag{1.10}.  The
assertions in~(b) are direct consequences of \theprotag{1.7(b)} {Theorem}.
For~(c) we use \theprotag{1.7(c)} {Theorem} and the fact that for any
disjoint union
  $$ \fld{M_1\sqcup M_2} = \fld{M_1}\times \fld{M_2}. $$
It remains to prove the gluing law~(d).  Fix a bundle $Q'\to\bX$.  Then for
each $Q\to Y$ and each $P\cut\in \fld{X\cut}(Q'\sqcup Q\sqcup Q)$ we have
  $$ \eac X{g_Q(P\cut)} = \Tr_Q\( \eac{X\cut}{P\cut}\) $$
by~\thetag{1.15}.  Fix a set of representatives~$\{Q\}$ for~$\fldb Y$.  Let
$\fldb X(Q')_Q$~denote the equivalence classes of bundles over~$X$ whose
restriction to~$\bX$ is~$Q'$ and to ~$Y$ is~$Q$.  Recall the gluing
map~\thetag{2.3} and the equation~\thetag{2.5} relating the measures.  Then
  $$ \spreadlines{6pt}
     \split
     Z_X(Q') &= \int_{\fldb X(Q')} d\mu \,(\Pb)\, \eac X{\Pb}\\
     &= \sum\limits_{Q\in \{Q\}}\int_{\fldb X(Q')_Q}d\mu\, (\Pb) \,\eac
     X{\Pb}\\
     &= \sum\limits_{Q\in \{Q\}}\int_{\fldb {X\cut}(Q'\sqcup Q\sqcup Q)}d\mu
     ([P\cut]) \,\mu _Q\, \Tr_Q\(\eac{X\cut}{[P\cut]}\)\\
     &= \sum\limits_{Q\in \{Q\}}\mu _Q\,\Tr_Q\bigl( Z_{X\cut}(Q'\sqcup
     Q\sqcup Q) \bigr).\endsplit  \tag{2.18}$$
The definition~\thetag{2.11} of the inner product shows that this is
equivalent to ~\thetag{2.17}.
\enddemo

\newpage
\head
\S{3} Surfaces with Boundary
\endhead
\comment
lasteqno 3@ 32
\endcomment

Now we specialize to~$d=2$---that is, to the $2+1$~dimensional theory---and
examine the more detailed structure associated to surfaces with boundary.
The classical theory of~\S{1} assigns a metrized complex line~$L_Q$ to each
$\Gamma $~bundle $Q\to Y$ over a closed oriented 2-manifold.  In this section
we construct lines when $Y$~has a boundary, but only after fixing certain
choices on the boundary.\footnote{The dependence on these choices leads to a
consideration of {\it gerbes\/}~\cite{B}, ~\cite{BM}, \cite{F5}.  We hope to
develop this idea in the quantum theory elsewhere.} These lines obey a gluing
law, which we state in \theprotag{3.2} {Theorem}.  The basepoints and
boundary parametrizations which appear in that theorem are part of the
process of fixing choices on the boundary.  By gluing cylinders these lines
lead to certain central extensions of subgroups of~$\Gamma $ \thetag{3.13},
which fit together into a central extension of a certain
groupoid~\thetag{3.9}.  The quantization then extends the definition
of~$E(Y)$ to surfaces with boundary and provides a method for computing this
vector space by cutting and gluing.  There is a rich algebraic structure: The
vector space attached to the cylinder is a {\it coalgebra\/}, each boundary
component of a surface determines a {\it comodule\/} structure on its quantum
space, and the gluing law appears naturally in terms of {\it cotensor
products\/}.\footnote{A more elaborate development of the formal properties
gives a tensor product description, see \cite{Q2,\P\P8,9} for a treatment of
finite gauge groups and an axiomatization.} An important point is the
behavior of the inner product under gluing~\thetag{3.26}, which comes quite
naturally in our approach.  The properties of these quantum Hilbert spaces
are stated in \theprotag{3.21} {Theorem}, which is the main result in this
section.  We analyze the algebraic structures more closely in~\S{4}.
Throughout this section we work with a fixed cocycle~$\ah\in C^3(B\Gamma
;\RZ)$.  We remark that there are several simplifications if~$\ah=0$, some of
which we discuss in~\S{5}.

We begin with the classical theory.  The important point is to rigidify the
data over the boundary of a surface.  Hence fix the standard circle
$S^1=[0,1]\bigm/0\sim 1$.\footnote{For the integration theory
of Appendix~B we also need to fix a standard cycle~$s\in C_1(\cir)$
which represents the fundamental class.} Consider first $\Gamma $~bundles
$R\to\cir$ with a basepoint in~$R$ chosen over the basepoint in~$\cir$.
Morphisms are required to preserve the basepoints.  We denote the category of
these pointed bundles and morphisms by~$\bfld {\cir}$.  Notice that there are
no nontrivial automorphisms, since a deck transformation which is the
identity at one point is the identity everywhere (on any connected space).
Further, the basepoint determines a holonomy map $\bfld{\cir}\to\Gamma $, and
the induced map $\bfldb{\cir}\to\Gamma $ on equivalence classes is a
bijection.  Summarizing, if $R_1,R_2$~are $\Gamma $~bundles over~$\cir$ with
the same holonomy, then there is a {\it unique\/} isomorphism $R_1\to R_2$
which preserves basepoints.

For each~$g\in \Gamma $ fix once and for all a pointed bundle ~$\Rb_g\to\cir$
with holonomy~$g$ and a classifying map
 $$ \phi _g\:\Rb_g\to E\Gamma  \tag{3.1}$$

Let $Y$~be a compact oriented 2-manifold.  Fix a diffeomorphism
$\cir\to(\partial Y)_i$ for each component~$(\partial Y)_i$ of~$\partial Y$.
A boundary component is labeled~`$+$' if the parametrization preserves
orientation and `$-$'~ otherwise.  The images of the basepoint in~$\cir$ give
a basepoint on each component of~$\partial Y$.  Then the $\Gamma $~bundles
$Q\to Y$ with basepoints chosen over the basepoints of~$\partial Y$ form a
category~$\bfld Y$; morphisms in this category are required to preserve the
basepoints.

     \proclaim{\protag{3.2} {Theorem}}
 Let $Y$~be a compact oriented 2-manifold with parametrized boundary.  Then
there is a functor
  $$ Q\longmapsto L_Q,\qquad Q\in\bfld Y, $$
which attaches to each $\Gamma $ ~bundle $Q\to Y$ with basepoints a metrized
line~$L_Q$.  It generalizes the corresponding functor~\thetag{1.8} for closed
surfaces, and satisfies the functoriality\footnote{The functoriality holds
for maps which preserve the basepoints and the boundary
parametrizations.}~\thetag{1.9}, orientation~\thetag{1.11}, and
additivity~\thetag{1.13} properties.  In addition it satisfies:\newline
 \hphantom{\rom(d\rom)}\rom({\it Gluing\/}\rom)\ Suppose $S\hookrightarrow Y$
is a closed oriented codimension one submanifold and $Y\cut$~the manifold
obtained by cutting along~$S$.  Then $\partial Y\cut = \partial Y\sqcup
S\sqcup -S$ and we use parametrizations which agree on~$S$ and~$-S$.  Suppose
$Q\in \bfld Y$~is a bundle over~$Y$ and $Q \cut\in \bfld{Y\cut}$~the induced
bundle over~$Y\cut$.  (We choose basepoints over~$S$ and~$-S$ which agree.)
Then there is a natural isometry
   $$ L_Q \cong L_{Q\cut}.   \tag{3.3}$$
     \endproclaim

     \demo{Proof}
 Let $Q\to Y$ be a pointed bundle.  Using the boundary parametrizations we
identity~$\partial Y$ as a disjoint union of circles.  Since pointed bundles
over the circle have no automorphisms, there is a unique based isomorphism
of~$\partial Q$ with a disjoint union of our standard pointed
bundles~$\Rb_g$.  Then the~$\phi _g$ chosen above~\thetag{3.1} determine a
classifying map $\phi \:\partial Q\to E\Gamma $.  Let $\Cal{C}_Q$ denote the
category of classifying maps~$f\:Q\to E\Gamma$ which extend~$\phi $.  A
morphism $f @>h>> f'$ is a homotopy~$h$ which is constant on~$\partial Q$, or
better a homotopy class rel boundary of such homotopies.  As in~\thetag{1.4}
we define a functor $\Cal{F}_Q\:\Cal{C}_Q \to\Cal{L}$ by
  $$ \Cal{F}_Q (f) = I_{Y,\bar{f}^*\ah} $$
and
  $$ \Cal{F}_Q (f @>h>> f') = e^{-2\pi iS_{S^1\times Y}(\cir\times Q)}
     \exp \left\{   2\pi i  \int_{\zo\times Y} \bar{h}^*\ah \right\} \:
     I_{Y,\bar{f}^*\ah} \longrightarrow I_{Y,\bar{f}'{}^*\ah}, \tag{3.4}$$
but now we must reinterpret the formul\ae.  First, the integration lines are
defined for manifolds with boundary in \theprotag{B.5} {Proposition} of
Appendix~B.  If $\partial Y=\emptyset $, then $S_{S^1\times Y}(\cir\times
Q)=0$ by the argument in~\thetag{1.5}.  If $\partial Y\not= \emptyset $, then
the prefactor in~\thetag{3.4} is an element in the line~$L_{\cir\times
\partial Q}$, which is trivialized by the classifying map~$\phi $ and the
trivialization~\thetag{B.9}.  (This follows from the construction
of~$L_{\cir\times \partial Q}$ in~\S{1}.  The integration line~\thetag{1.3}
in that construction is trivialized using the parametrization of~$\partial Y$
and~\thetag{B.9}.)  So this prefactor is a complex number.  Since
  $$ \partial (\zo\times Y) = \{1\}\times Y \,\cup\, -\{0\}\times Y\ \,\cup\,
     -\zo\times \partial Y, $$
the second factor in~\thetag{3.4} is an element of
  $$ I_{Y,\bar{f}'{}^*\ah} \otimes I_{Y,\bar{f}{}^*\ah}^* \otimes
     I^*_{\zo\times \partial Y,\phi ^*\ah}. \tag{3.5}$$
Here we use \theprotag{B.5} {Proposition}, in particular the gluing
law~\thetag{B.6}.)  But the boundary parametrization and the
trivialization~\thetag{B.8} trivialize the last factor in~\thetag{3.5}.
Hence \thetag{3.4}~is well-defined.  The fact that $\Cal{F}_Q$~defines a
functor and that this functor has no holonomy are routine checks.  Both use
the compatibility of~\thetag{B.8} and~\thetag{B.9} under gluing.
Define~$L_Q$ to be the line of invariant sections of~$\Cal{F}_Q$.

We leave the verification of~\thetag{3.3} to the reader.
     \enddemo

As a corollary we obtain a metrized line bundle
  $$ \Cal{L}_Y \longrightarrow \bfld Y  $$
which generalizes~\thetag{1.19}.  Furthermore, we remark that
\thetag{2.1}~defines a measure on~$\bfld Y$ which is invariant under
morphisms, so passes to a measure on the set of equivalence classes~$\bfldb
Y$.  If each component of~$Y$ has nonempty boundary, then this measure has
unit mass on each bundle.

The fact that pointed bundles over~$\cir$ have no automorphisms makes gluing
well-defined on {\it equivalence classes\/} of bundles.  More precisely,
consider a compact oriented 2-manifold~$Y$ with parametrized boundary, an
oriented codimension one submanifold $S\hookrightarrow Y$, and the resulting
cut manifold~$Y\cut$.  Suppose $Q\cut\in \bfld{Y\cut}$ restricts to
isomorphic bundles over the two copies of~$S$ in~$Y\cut$.  Then there is a
well-determined bundle $g(Q\cut)\in \bfld Y$ obtained by gluing.
(Compare~\thetag{2.3}.)  In other words, there is a gluing map
  $$ g\:\Cal{B}^{\vphantom{\prime}}_{Y\cut}\subset \bfld{Y\cut}
     \longrightarrow \bfld Y $$
defined on the subset~$\Cal{B}^{\vphantom{\prime}}_{Y\cut}$ of bundles which
are isomorphic on the two copies of~$S$.  Since the isomorphism used to
perform the gluing is unique, there is an induced gluing
  $$ g\: \overline{\Cal{B}^{\vphantom{\prime}}_{Y\cut}} \subset
     \bfldb{Y\cut}\longrightarrow \bfldb Y \tag{3.6}$$
on equivalence classes of bundles.  Suppose $Y\cut = Y_1 \sqcup Y_2$ and
$Q\cut = Q_1 \sqcup Q_2$.  Further, suppose the copy of~$S$ in~$Y_1$ has a
$+$~parametrization and the copy in~$Y_2$ has a $-$~parametrization.  Then we
denote the glued bundle over~$Y$ as
  $$ Q_1 \circ _S Q_2 = Q_1\circ Q_2. $$
\theprotag{3.2(d)} {Theorem} implies that the gluing lifts to the associated
lines.  In other words, there is an isometry
  $$ L_{Q_1}\otimes L_{Q_2}\longrightarrow L_{Q_1\circ Q_2}. \tag{3.7}$$
As gluing makes sense on the equivalence classes, we write $\Qb_1\circ \Qb_2$
for the glued element in~$\bfldb Y$.  Then \thetag{3.7}~induces an isometry
  $$ L_{\Qb_1}\otimes L_{\Qb_2}\longrightarrow L_{\Qb_1\circ \Qb_2}.
     \tag{3.8}$$
(Recall, however, that these ``lines'' may be the zero vector space.)

We apply this first to the cylinder~$Y=\zo\times \cir$ which we cut along
$\{1/2\} \times \cir$.  Then $\bfldb{Y\cut} = \bfldb Y\times \bfldb Y$, and
the gluing provides a {\it groupoid\/} structure on
  $$ \Cal{G}=\bfldb{\zo\times\cir}.   \tag{3.9}$$
Let $\{*\}\in \cir$ be the basepoint.  An element\footnote{We use `$\bcyl $'
to denote bundles over the cylinder and `$Q$'~to denote bundles over
arbitrary surfaces.}~$\bbcyl_{\langle x,g\rangle }\in \Cal{G}$ is given by a
pair $\langle x,g\rangle \in \Gamma \times \Gamma $, where $x$~is the
holonomy around~$\{0\}\times \cir$ and $g$~is the parallel transport along
{}~$\zo\times \{*\}$.  This parallel transport is well-defined since the bundle
has a basepoint over~$\{0\}\times \{*\}$ and one over~$\{1\}\times \{*\}$.
The groupoid composition is then
  $$ \bbcyl_{\langle x_1,g_1\rangle }\circ \bbcyl_{\langle x_2,g_2\rangle } =
     \bbcyl_{\langle x_1,g_2g_1\rangle },\qquad \text{if $x^{\vphantom{1}}_2
     = g^{\vphantom{1}}_1 x^{\vphantom{1}}_1g_1\inv $},  \tag{3.10}$$
and is undefined if~$x^{\vphantom{1}}_2\not=
g^{\vphantom{1}}_1x^{\vphantom{1}}_1 g_1\inv $.  Let $L_{\langle x,g\rangle }
= L_{\bbcyl_{\langle x,g\rangle }}$ denote the line attached to the
equivalence class~$\bbcyl_{\langle x,g\rangle }$.  Note that $\dim L_{\langle
x,g\rangle } =1$ since pointed bundles over the cylinder have trivial
automorphism groups.  Then \thetag{3.7}~in this context is an isometry
  $$ L_{\langle x_1,g_1\rangle }\otimes L_{\langle x_2,g_2\rangle
     }\longrightarrow L_{\langle x_1,g_2g_1\rangle },\qquad \text{if
     $x^{\vphantom{1}}_2 = g^{\vphantom{1}}_1x^{\vphantom{1}}_1g_1\inv $.}
     \tag{3.11}$$
In particular, for $g_1=g_2=e$ and any~$x\in \Gamma $ this gives a
trivialization
  $$ L_{\langle x,e\rangle }\cong \CC. \tag{3.12}$$
Restricting to unit vectors the isometries~\thetag{3.11} define a central
extension~$\hat{\Cal{G}}$ of the groupoid~$\Cal{G}$ by~$\TT$.  For each $x\in
\Gamma $ the set
  $$ \{\bbcyl_{\langle x,g\rangle } : gx=xg\} $$
is closed under the composition~\thetag{3.10} and is isomorphic\footnote{In
fact, it is anti-isomorphic to~$C_x$ because the composition
law~\thetag{3.10} reverses the order of multiplication.} to the
centralizer~$C_x$ of~$x$ in~$\Gamma $.  The lines~\thetag{3.11} then give a
central extension
  $$ 1 @>>> \TT @>>> \hat{C}_x @>>> C_x @>>> 1. \tag{3.13}$$
The equivalence class of this extension can be expressed in terms of the
cohomology class $[\ah]\in H^3(B\Gamma ;\RZ)$.  Namely, there is a homotopy
equivalence
  $$ \Map(\cir,B\Gamma ) \sim \coprod_{[x]}BC_x, $$
where $x$~runs over representatives of the conjugacy classes in~$\Gamma $.
Let
  $$ e\:\cir\times \Map(\cir,B\Gamma )\longrightarrow B\Gamma  $$
be the evaluation map and
  $$ \pi \:\cir\times \Map(\cir,B\Gamma )\longrightarrow \Map(\cir,B\Gamma )
     $$
the projection.

     \proclaim{\protag{3.14} {Proposition}}
 The cohomology class of the extension~\thetag{3.13} is (a component of) the
transgression
   $$ \pi _*e^*[\ah] \in \bigoplus_{[x]} H^2(BC_x;\RZ). $$
     \endproclaim

     \demo{Proof}
 First, we recall that the central extension~\thetag{3.13} determines a class
in $H^2(C_x;\RZ)$, the second group cohomology.  To construct a cocycle, for
each~$g\in C_x$ choose an element $\gh\in \hat{C}_x$ covering~$g$.  Then set
  $$ c(g^{\vphantom{1}}_1,g^{\vphantom{1}}_2) =
     \widehat{g^{\vphantom{1}}_1g^{\vphantom{1}}_2}\gh_2\inv \gh_1\inv
     \subset \RZ,\qquad g_1,g_2\in C_x. \tag{3.15}$$
This is a cocycle for the group cohomology.

Next, the group cohomology is isomorphic to the (singular) cohomology of the
classifying space as follows.  Let $\hat{\beta} \in C^2(BC_x;\RZ)$ be a
singular cocycle.  Then for each $g\in BC_x$ fix a based loop $\gamma
_g\:\cir\to BC_x$ whose homotopy class corresponds to~$g$ under the
isomorphism $\pi _1(BC_x) \cong C_x$.  Denote $I_g = I_{\cir,\gamma
_g^*\hat{\beta}}$ for the integration line of \theprotag{B.1} {Proposition}.
Choose an element $\epsilon _g\in I_g$ of unit norm.  For~$g_1,g_2\in
C_x$ the composite loop~$\gamma _{g_1}\gamma _{g_2}$ is homotopic to~$\gamma
_{g_1g_2}$.  Let
  $$ k_{g_1,g_2}\:\zo\times \cir\longrightarrow BC_x  $$
be a homotopy.  By the integration theory of Appendix~B this determines an
isometry\footnote{We use the $d=1$~case of \thetag{B.6} to identify the
integration line of~$\gamma _{g_1}\gamma _{g_2}$ with $I_{g_1}\otimes
I_{g_2}$ (cf.~the footnote preceding \theprotag{B.5} {Proposition}).  Also,
by~\thetag{B.3} the map~$\theta _{g_1,g_2}$ only depends on the homotopy
class of~$k_{g_1,g_2}$.}
  $$ \theta _{g_1,g_2} = \int_{\zo\times \cir}k_{g_1,g_2}^* \hat{\beta} \:
     I_{g_1}\otimes I_{g_2}\longrightarrow I_{g_1g_2}. \tag{3.16}$$
Define $c'(g_1,g_2)\in \RZ$ by
  $$ e^{\tpi c'(g_1,g_2)}\theta _{g_1,g_2}(\epsilon _{g_1}\otimes \epsilon
     _{g_2}) = \epsilon _{g_1g_2}. \tag{3.17}$$
An easy check shows that $c'$~is a cocycle for group cohomology.

With these preliminaries aside we proceed to the proof.  For each~$g\in C_x$
fix a pointed bundle $T_{\langle x,g \rangle}\to \zo\times \cir$ with the
correct holonomy.  By gluing together the two ends of the cylinder, we obtain
a bundle $\check T_{\langle x,g \rangle}\to\cir\times \cir$.  By~\thetag{3.3}
the lines corresponding to~$T_{\langle x,g \rangle}$ and $\check T_{\langle
x,g \rangle}$ are isomorphic, so we pass freely between them.  Choose a
classifying map $f_{\langle x,g \rangle}\:T_{\langle x,g \rangle}\to E\Gamma
$ which restricts to~$\phi _x$ on each boundary component.  Let $\check
f_{\langle x,g \rangle}\:\check T_{\langle x,g \rangle}\to E\Gamma $ be the
induced classifying map.  By~\thetag{1.3}, $\check f_{\langle x,g
\rangle}$~determines an isometry
  $$ L_{\langle x,g \rangle}=L_{\check T_{\langle x,g \rangle}}\cong
     I_{\cir\times \cir,\bar{f}_{\langle x,g  \rangle}^*\ah} . $$
Denote this line as ~$I_g$ and fix an element $\epsilon _g\in I_g$ of unit
norm.  From the point of view of the central extension~\thetag{3.13}, it is
an element ~$\gh\in \hat{C}_x$ which lifts~$g\in C_x$.  (Recall that
$\hat{C}_x$~is defined as the set of elements of unit norm in the
lines~$L_{\langle x,g \rangle}$.)  For each $g_1,g_2\in C_x$ we have two
trivializations of~$L_{\langle x,g_1g_2 \rangle}$, via~\thetag{3.11}; their
ratio is a cocycle~$c(g_1,g_2)$ of the central extension~\thetag{3.15}.  On
the other hand, their ratio may be computed from~\thetag{1.4}.  Namely,
choose a homotopy
  $$ h_{g_1,g_2}\: \zo\times S^1\times \cir\longmapsto E\Gamma  $$
from the ``composite''~$f_{\langle g\mstrut _1xg_1\inv ,g\mstrut _2
\rangle}f_{\langle x,g\mstrut _1 \rangle}$ (computed by gluing the first map
over~$\{1\}\times \cir$ to the second map over~$\{0\}\times \cir$ and
rescaling) to~$f_{\langle x,g_1g_2 \rangle}$.  Then
  $$ \theta _{g_1,g_2} = \int_{\zo\times \cir\times \cir}
     \bar{h}_{g_1,g_2}^*\ah\: I_{g_1}\otimes I_{g_2}\longrightarrow
     I_{g_1g_2},  \tag{3.18}$$
and the desired cocycle $c(g_1,g_2)\in \RZ$ is determined by~\thetag{3.17} as
before.
Let
  $$ \bar{k}_{g_1,g_2}\:\zo\times \cir\longrightarrow \Map(\cir,B\Gamma )
     $$
be the map~$\bar{h}_{g_1,g_2}$ as a function of its first two variables.
Then \thetag{3.18} implies
  $$ \theta _{g_1,g_2} = \int_{\zo\times \cir}\bar{k}_{g_1,g_2}^*\pi
     _*e^*(\ah). \tag{3.19}$$
A comparison of~\thetag{3.19} and~\thetag{3.16} shows that~$c$, which is
defined as a cocycle in the group cohomology for the central extension,
corresponds to a cocycle for the cohomology class~$\pi _*e^*[\ah]$, as
claimed.
     \enddemo

Finally, we remark that the measure~\thetag{2.1} assigns unit mass to each
$\bbcyl_{\langle x,g\rangle }\in \Cal{G}$, so is obviously left and right
invariant under the groupoid composition law.

Now suppose $Y$~is a compact oriented 2-manifold with parametrized boundary.
Suppose $(\partial Y)_i$~is a $+$~boundary component.  Then if $\Qb\in \bfldb
Y$ and $\bbcyl\in \Cal{G}$ agree on $(\partial Y)_i\subset Y$ and $\{0\}
\times \cir\subset \zo\times \cir$ there is a glued bundle $\Qb\circ
\bbcyl\in \bfldb Y$.  Thus a $+$~boundary component determines a right
$\Cal{G}$~action on~$\bfldb Y$.  The isometry~\thetag{3.8}
  $$ L_{\Qb}\otimes L_{\bbcyl}\longrightarrow L_{\Qb\circ \bbcyl}
     \tag{3.20}$$
is a lift to a right $\hat{\Cal{G}}$~action on an extension of~$\bfldb Y$.
Similarly, a $-$~boundary component determines a left action.  It is easy to
see that these actions preserve the measure on~$\bfldb Y$.

The quantization generalizes ~\thetag{2.10}.  Namely, for each compact
oriented 2-manifold~$Y$ with parametrized boundary, set
  $$ E(Y) = L^2 (\bfldbu Y,\overline{\Cal{L}\mstrut _Y}). $$
The preceding classical structure has quantum implications.  To state these
we need the following algebraic notions.  (See Appendix~A for a more detailed
review of coalgebras and comodules.)  Suppose $A$~is a coalgebra over~$\CC$,
$E_R$~is a right $A$-comodule and $E_L$~is a left $A$-comodule.  Then the
{\it cotensor product\/}~$E_R \boxtimes_A E_L$ is the vector subspace of~$E_R
\otimes _{\CC}E_L$ annihilated by~$\Delta_R \otimes \id - \id \otimes \Delta
_L$, where $\Delta _R, \Delta _L$~are the coproducts.  If $E_R,E_L$~are
unitary comodules, then $E_R \boxtimes_A E_L$ inherits the subspace inner
product.  More generally, if $E$~has a left $A$-comodule structure~$\Delta
_L$ and a right $A$-comodule structure~$\Delta _R$, then we define
  $$ \Inv_A(E)\subset E $$
to be the subspace annihilated by~$\Delta _L - P\Delta _R$, where
$P\:E\otimes _\CC A \to A \otimes _{\CC}E$ is the natural isomorphism.

The following generalizes \theprotag{2.13} {Theorem}.

     \proclaim{\protag{3.21} {Theorem}}
 The assignment
  $$ Y\longmapsto E(Y) $$
of a hermitian vector space to a compact oriented 2-manifold with
parametrized boundary agrees with~\thetag{2.14} on closed manifolds and
satisfies:\newline
 \rom(a\rom)\ \rom({\it Functoriality\/}\rom)\ If $f\:Y'\to Y$ is an
orientation preserving
diffeomorphism which preserves the boundary parametrizations, then there is
an induced isometry
  $$ f_*\:E(Y') \longrightarrow E(Y)   $$
and these compose properly.\newline
 \rom(b\rom)\ \rom({\it Orientation\/}\rom)\ There is a natural isometry
  $$ E(-Y) \cong \overline{E(Y)}. \tag{3.22}$$
 \rom(c\rom)\ \rom({\it Multiplicativity\/}\rom)\ If $Y=Y_1\sqcup Y_2$ is a
disjoint union,
then there is a natural isometry
  $$ E(Y _1\sqcup Y _2) \cong E(Y _1)\otimes E(Y _2). \tag{3.23}$$
 \rom(d\rom)\ \rom({\it Coalgebra\/}\rom)\ Let $S$~be a parametrized closed
oriented 1-manifold.  Then
  $$ A_S = E(\zo\times S) $$
is a unitary\footnote{The notion of unitarity for coalgebras and comodules is
defined in~\thetag{A.3} and~\thetag{A.4}.} coalgebra with antiinvolution.
There are natural isomorphisms
  $$ A\mstrut _{-S}\cong A_S^{\text{op}}  \tag{3.24}$$
with the opposite coalgebra, and
  $$ A_{S_1\sqcup S_2}\cong A_{S_1}\times A_{S_2}. \tag{3.25}$$
 \rom(e\rom)\ \rom({\it Comodule\/}\rom)\ $E(Y)$~is a unitary right
$A_{\partial Y}$-comodule.  The isometries~\thetag{3.22} and~\thetag{3.23}
are compatible with the comodule structure.\newline
 \rom(f\rom)\ \rom({\it Gluing\/}\rom)\ Suppose $S\hookrightarrow Y$ is a
closed embedded
circle and $Y\cut$~the manifold obtained by cutting along~$S$.  So $\partial
Y\cut = \partial Y\sqcup S\sqcup -S$ and we use parametrizations which agree
on~$S$ and~$-S$.  Thus $E(Y\cut)$~is both a right and left $A_S$-comodule.
Then there is an isometry
  $$ E(Y) \cong \frac{1}{\#\Gamma }\cdot  \Inv_{A_S} E(Y\cut). \tag{3.26}$$
     \endproclaim

\flushpar
 Since all of our 1-manifolds are parametrized we can identify them with a
union of circles.  Let
  $$ A=A_{\cir}  \tag{3.27}$$
be the coalgebra attached to the standard circle.  Then \thetag{3.24}~ and
\thetag{3.25}~imply that any $A_S$~can be naturally identified with a direct
product of copies of~$A$ and~$A^{\text{op} }$.  Also, if $E$~is a hermitian
vector space and $\lambda $~a positive number, then $\lambda \cdot E$~denotes
the same underlying vector space with the hermitian product multiplied by a
factor of~$\lambda $.  Equation~\thetag{3.26} is the gluing law which allows
one to compute the vector space attached to a surface by cutting and pasting.
The isometry~\thetag{3.26} is compatible with~(a)--(c) and~(e).

     \demo{Proof}
 We only comment on some of the assertions.  As a warmup, consider the
situation where a finite group~$G$ acts on a finite set~$Z$.  The space
of functions~$\Cal{F}(G)$ is a coalgebra and $\Cal{F}(Z)$~is a comodule.
The coalgebra structure is dual to the multiplication~$G \times G
\to G $ and the comodule structure is dual to the action ~$Z\times
G \to Z$.  The counit is dual to the inclusion of the identity
element~$1\to G $.  If $G $~has a bi-invariant measure then
$L^2(G )$~has a compatible inner product.  If $Z$~has a measure
preserved by the $G $~action, then $L^2(Z)$~also has a compatible inner
product.

Our situation is different in two ways: We have a groupoid and we consider
functions with values in complex lines.  The latter is not significant, since
the lines form a central extension of the groupoid.  But the fact that we
only have a groupoid, and not a group, means that the coproduct of an element
is not naturally defined away from composable elements.  In ~\thetag{3.28} we
extend by zero.  This works.\footnote{In more complicated situations it is
not immediately clear what should be done.} Thus if $a\in A_S$ for some
parametrized compact oriented 1-manifold~$S$, then we define the coproduct
  $$ \Delta a\,(\bbcyln 1,\bbcyln 2) = \cases a(\bbcyln 1\circ \bbcyln 2)
     ,&\text{if $\bbcyln 1\circ \bbcyln 2$ is
     defined};\\0,&\text{otherwise}.\endcases   \tag{3.28}$$
for $\bbcyln 1,\bbcyln 2\in \bfldb{\zo\times S}$.  Note that this uses the
isometry~\thetag{3.11}.  The counit is
  $$ \epsilon (a) = \sum\limits_{x} a(\bbcyl_{\langle x,e\rangle }),
     \tag{3.29}$$
where we use the trivialization~\thetag{3.12}.  Similarly, if $Y$~is a
compact oriented 2-manifold with parametrized boundary, then the comodule
structure on~$E(Y)$ is defined by
  $$ \Delta x\,(\Qb,\bbcyl) = \cases x(\Qb\circ \bbcyl)
     ,&\text{if $\Qb\circ \bbcyl$ is
     defined};\\0,&\text{otherwise},\endcases $$
where $x\in E(Y)$, $\Qb\in \bfldb Y$, and $\bbcyl\in \bfldb{\zo\times
\partial Y}$.  Here we use the isometry~\thetag{3.20}.  The antiinvolution is
  $$ \bar{a}(\bbcyl) = \overline{a(\bbcyl\inv )}. $$
The compatibility of the $L^2$~inner product with these coproducts, expressed
by~\thetag{A.3} and~\thetag{A.4}, is a simple change of variables.  For
example, if $a,b,c\in A_S$, then
  $$ \split
     (\Delta a,b\otimes c) &= \sum\limits_{\bbcyln 1,\bbcyln 2} a(\bbcyln
     1\circ \bbcyln 2)\; \overline{b(\bbcyln 1)} \;\overline{c(\bbcyln 2)}\\
     &= \sum\limits_{\bbcyln 2,\bbcyln 3} a(\bbcyln 3) \;\overline{c(\bbcyln 2
     \inv)} \;\overline{b(\bbcyln 3\circ\bbcyln 2)}\\
     &= (a\otimes \bar{c},\Delta b).\endsplit $$

For the gluing law~(f), let $\overline{\Cal{B}_{Y\cut}} \subset
\bfldb{Y\cut}$ denote the subset of bundles which are isomorphic over the two
copies of~$S$.  We claim that any function $x\in \Inv_{A_S}E(Y\cut)$ has
support in~$\overline{\Cal{B}_{Y\cut}}$.  This follows since by definition
  $$ x(\Qbcut\circ _S\bbcyl) = x(\bbcyl\circ _{-S}\Qbcut) \tag{3.30}$$
for all $\Qbcut\in \bfldb{Y\cut}$ and~$\bbcyl\in \bfldb{\zo\times S}$, where
we understand the evaluation of~$x$ on an undefined composition to vanish.
Hence given any $\Qbcut\in \bfldb{Y\cut}$ we choose $\bbcyl = \Qbcut\res
{\zo\times S}$.  Then if $\Qbcut\res{-S} \not\cong \Qbcut\res S $, the
composition $\bbcyl\circ _{-S}\Qbcut$~ is undefined, whence
  $$ x(\Qbcut) = x(\Qbcut\circ \bbcyl) = 0 $$
as claimed.  Note that if $\Qbcut\res{-S} \cong  \Qbcut\res S$ then
\thetag{3.30}~is equivalent to
  $$ x(\bbcyl\inv \circ _{-S}\Qbcut\circ _S\bbcyl) = x(\Qbcut).
     $$

Next, consider the diagram
  $$ \CD
    \overline{\Cal{B}_{Y\cut}} @>g>> \bfldb Y\\
    @VrVV\\
    \bfldb S \endCD $$
where $g$~is the gluing map~\thetag{3.6} and $r$~is restriction.  From
\theprotag{2.4(a)} {Lemma} we conclude that $g$~is surjective.  Furthermore,
it follows from part~(b) of that lemma that if $\Qbcut\in g\inv (\Qb)$, then
any other element of~$g\inv (\Qb)$ is of the form $\bbcyl\inv \circ
_{-S}\Qbcut\circ _S\bbcyl$, where $\bbcyl\in \bfldb{\zo\times S}$ restricts
over~$\{0\}\times S$ to a bundle isomorphic to~$\Qbcut\res S$.  Now
\theprotag{3.2(d)} {Theorem} implies that the pullback
of~$\overline{\Cal{L}_Y}$ via~$g$ is isomorphic to the restriction
of~$\overline{\Cal{L}_{Y\cut}}$ to~$\overline{\Cal{B}_{Y\cut}}$.  It now
follows that pullback via~$g$ is an isomorphism $E(Y)\cong
\Inv_{S_S}E(Y\cut)$.

It remains to compare the inner products.  We claim that for any~$\Qb\in
\fldb Y$,
  $$ \mu _{\Qb} = \frac{\voll\bigl(g\inv (\Qb)\bigl)}{\#\Gamma }.
     \tag{3.31}$$
For in our situation \thetag{2.5}~becomes
  $$ \mu _{\Qb} = \frac{\voll\bigl(g\inv (\Qb)\bigl)}{\#r\bigl(g\inv
     (\Qb)\bigr)\cdot \#\Aut(\Qb\res S) }. \tag{3.32}$$
Note that $\Qb\res S$~is only determined up to isomorphism and has no
basepoint; nevertheless the number $\#\Aut(\Qb\res S)$~is independent of the
representative chosen.  The factor~$\#r\bigl(g\inv (\Qb)\bigr)$ counts these
representatives.  Equation~\thetag{3.31} follows immediately
from~\thetag{3.32} since $\Gamma $~acts transitively on these representatives
(changing the basepoint) with stabilizer the automorphism group.  Now
consider $x,y\in L^2(\bfldbu Y,\overline{\Cal{L}\mstrut _Y})$.  Then
  $$ \spreadlines{6pt}
     \aligned
     (g^*x,g^*y)_{E(Y\cut)} &= \int_{\overline{\Cal{B}_{Y\cut}}} d\mu
     (\Qbcut)\, x\bigl(g(\Qbcut)\bigr)\,\overline{y\bigl(g(\Qbcut)\bigr)}\\
     &= \int_{\fldb Y}d\mu (\Qb)\,\#\Gamma \cdot x(\Qb)\,\overline{y(\Qb)}\\
     &= \#\Gamma \cdot (x,y)_{E(Y)}.\endaligned $$
This completes the proof of~\thetag{3.26}
     \enddemo

\newpage
\head
\S{4} The Modular Functor
\endhead
\comment
lasteqno 4@ 16
\endcomment

Our first task in this section is to construct a {\it modular
functor\/}~\cite{S1} from the quantum Hilbert spaces of \theprotag{3.21}
{Theorem}.\footnote{The modular functor encodes the structure of {\it
conformal blocks\/} in rational conformal field theory~\cite{MS}.} We first
dualize the coalgebras and comodules of that theorem to obtain the more
familiar algebras and modules, and so we make contact with standard structure
theory of these objects.  (It is also possible to develop an algebra version
directly~\cite{Q1,\S4}.)  A key point is semisimplicity (\theprotag{4.2}
{Proposition}) which leads to a set of {\it labels\/}.  Then from the
comodule~$E(Y)$ for $Y$~an oriented surface with boundary we derive the
vector space~\thetag{4.7} typically attached to a surface with labeled
boundary.  We also recover the {\it Verlinde algebra\/} (\theprotag{4.14}
{Proposition}).  It is important to notice that not only does $E(Y)$~have a
richer structure than the vector spaces attached to the surfaces with labeled
boundary, but the inner product structure is more natural there.  (Compare
the gluing laws~\thetag{3.26} and~\thetag{4.11}.)

We first analyze the coalgebra~$A$ \thetag{3.27} in more detail.  Since
algebras are more familiar than coalgebras, we switch to the dual
algebra~$A^*$.  We have
  $$ A^* = \bigoplus _{\bbcyl} L^*_{\bbcyl}, \tag{4.1}$$
where $\bbcyl$~ranges over~$\Cal{G}$.  If $\xi _1\in L^*_{\bbcyln 1}$ and
$\xi _2\in L^*_{\bbcyln 2}$, then $\xi_1 \xi _2\in L^*_{\bbcyln 1\circ
\bbcyln2}$ is obtained from~\thetag{3.11} if $\bbcyln1\circ \bbcyln2$~exists
and is zero if $\bbcyln1$~and $\bbcyln2$~are not composable.

     \proclaim{\protag{4.2} {Proposition}}
 The algebra ~$A^*$ is semisimple.  Hence
  $$ A^*\cong \prod\limits_{\lambda \in \Phi }M_\lambda \tag{4.3}$$
is isomorphic to a direct product of matrix algebras.
     \endproclaim

\flushpar
 The proof imitates the standard proof of Maschke's theorem in the theory of
finite group representations.

     \demo{Proof}
 It suffices to show that every $A^*$-module is completely reducible, or
equivalently that any surjective morphism of $A^*$-modules
  $$ P_1 @>f>> P_2 \tag{4.4}$$
splits.  For this suppose that $P_2@>g>>P_1$ is a $\CC$-linear splitting
of~$f$.  For each $\bbcyl\in \Cal{G}$ fix a nonzero element $\xi _{\bbcyl}\in
L^*_{\bbcyl}$ and consider
  $$ g' = \frac{1}{\# \Cal{G}}\sum\limits_{\bbcyl}\xi \mstrut _{\bbcyl}g\xi
     \inv _{\bbcyl}, $$
where $\xi \inv _{\bbcyl}\in L^*_{\bbcyli}$ is the unique element such that
$\xi \inv _{\bbcyl}\otimes \xi \mstrut _{\bbcyl}\in L_{\bbcyli}^*\otimes
L_{\bbcyl}^*\cong \CC$ is the unit element (cf\.~\thetag{3.12}).  Since $g$~is
$\CC$-linear, the map~$g'$ is independent of the choice of~$\xi _{\bbcyl}$.
We claim that $g'$~is an $A^*$-module homomorphism which
splits~\thetag{4.4}.  The calculation that $g'f=\id_{P_1}$ is
straightforward.  To see that $g'$~commutes with the $A^*$-action, suppose
that $\xi _0\in L^*_{\bbcyln0}$.  Set
  $$ \xi '_{\bbcylp} = \xi _0\inv \xi \mstrut _{\bbcyl},\qquad \bbcylp =
     \bbcyln0\inv \circ \bbcyl, $$
when the composition makes sense.  Only these compositions appear in the
calculation
  $$ \aligned
     g'\xi _0 &= \frac{1}{\#\Cal{G}}\sum\limits_{\bbcyl}\xi \mstrut
     _{\bbcyl}g\xi \inv _{\bbcyl}\xi \mstrut _0 \\
     &= \frac{1}{\#\Cal{G}}\sum\limits_{\bbcylp}\xi \mstrut _0\xi
     '_{\bbcylp}g\xi \inv _{\bbcyl'} \\
     &= \xi _0 g',\endaligned $$
which completes the proof.
     \enddemo

Let $\Phi =\{\lambda \}$~be the set of {\it labels\/} in~\thetag{4.3}.  This
set labels the irreducible representations of~$A^*$, or equivalently the
irreducible corepresentations~$E_{\lambda }$ of~$A$.  Fix a hermitian
structure on~$E_\lambda $ compatible with the $A$~action; it is unique up to
a scalar multiple.  The set~$\Phi $ also labels the irreducible
corepresentations of~$A\op$; then the label~$\lambda $ corresponds
to~$E_\lambda ^*$.  Now suppose $Y$~is a compact oriented 2-manifold with
parametrized boundary.  Then using \theprotag{A.7} {Proposition} we can split
the unitary comodule~$E(Y)$ according to the irreducible corepresentations
of~$A_{\partial Y}\cong A\times \cdots \times A\times A\op\times \cdots
\times A\op$.  In other words, there is an isometry
  $$ E(Y)\cong \bigoplus_{\bal} E(Y,\bal)\otimes E_{\bal}, \tag{4.5}$$
where $\bal=\langle \lambda _1,\dots ,\lambda _k\rangle $ is a labeling of the
boundary components,
  $$ E_{\bal} = E^{\pm 1}_{\lambda _1}\otimes \dots \otimes E_{\lambda
     _k}^{\pm1}  \tag{4.6}$$
is the corresponding corepresentation of~$A_{\partial Y}$, and
  $$ E(Y,\bal) = \Hom_{A_{\partial Y}}\bigl( \frac{E_{\bal}}{\dim
     E_{\bal}},E(Y)\bigr). \tag{4.7}$$
We use~$E_\lambda $ in~\thetag{4.6} if the corresponding boundary component
is positive, and we use the dual~$E_\lambda \inv =E_\lambda ^*$ if the
corresponding boundary component is negative.  So $E(Y,\bal)$~is the inner
product space attached to a compact oriented surface with parametrized {\it
labeled\/} boundary.  The assignment
  $$ \langle Y,\bal\rangle \longmapsto E(Y,\bal) \tag{4.8}$$
is termed a (reduced) {\it modular functor\/} by Graeme Segal~\cite{S1}; in
the physics literature $E(Y,\bal)$~is called a {\it conformal block\/}.
\theprotag{3.21} {Theorem} easily implies the following properties of these
inner product spaces.

     \proclaim{\protag{4.9} {Proposition}}
 The assignment~\thetag{4.8} satisfies:\newline
 \rom(a\rom)\ \rom({\it Functoriality\/}\rom)\ If $f\:\langle Y',\bal'\rangle
\to \langle
Y,\bal\rangle $ is an orientation preserving diffeomorphism which preserves
the boundary parametrizations and the labels, then there is an induced
isometry
  $$ f_*\:E(Y',\bal') \longrightarrow E(Y,\bal)  $$
and these compose properly.\newline
 \rom(b\rom)\ \rom({\it Orientation\/}\rom)\ There is a natural isometry
  $$ E(-Y,\bal) \cong \overline{E(Y,\bal)}.  $$
 \rom(c\rom)\ \rom({\it Multiplicativity\/}\rom)\ If $\langle Y,\bal\rangle =
\langle
Y_1,\bal_1\rangle \sqcup \langle Y_2,\bal_2\rangle $ is a disjoint union,
then there is a natural isometry
  $$ E(Y _1\sqcup Y _2,\bal_1\cup\bal_2) \cong E(Y _1,\bal_1)\otimes E(Y
  _2,\bal_2).   \tag{4.10}$$
 \rom(d\rom)\ \rom({\it Gluing\/}\rom)\ Suppose $Y\cut$~is the manifold
obtained from~$Y$ by cutting along an embedded circle~$S$.  Let $\bal$~be a
labeling of~$\partial Y$.  Then there is an isometry
  $$ E(Y,\bal) \cong \bigoplus_{\mu \in \Phi }\frac{\dim E_\mu
     }{\#\Gamma }\cdot E(Y\cut,\bal\cup\mu \cup\mu ).  \tag{4.11}$$
     \endproclaim

\flushpar
 The extra factor in the gluing law~\thetag{4.11} does not appear in Segal's
work~\cite{S1}, but it does appear in Walker's treatment~\cite{Wa} of the
$SU(2)$~Chern-Simons theory.  We should point out that the inner product
in~$E(Y,\bal)$ can be scaled, as can the isomorphism in~\thetag{4.11}.  There
are choices that would eliminate the extra factor in the gluing law.  Our
scale choices seem quite natural, nonetheless.  The isometry~\thetag{4.11} is
independent of the choice of hermitian structure on~$E_\lambda $ (which is
unique up to scale).

     \demo{Proof}
 Equation~\thetag{4.10} follows immediately from~\thetag{3.23}.
For~\thetag{4.11} we rewrite~\thetag{3.26} using~\thetag{4.5}
and~\thetag{A.8}:
  $$ \aligned
     E(Y)&\cong \bigoplus_{\bal} E(Y,\bal)\otimes E_{\bal}\\
     &\cong \frac{1}{\#\Gamma } \bigoplus_{\bal,\mu ,\nu }
     E(Y\cut,\bal\cup\mu \cup\nu ) \otimes E_{\bal} \otimes \Inv _{A_S}
     (E_\mu \otimes E^*_\nu ) \\
     &\cong \bigoplus_{\bal,\mu }\frac{\dim E_\mu }{\#\Gamma } \cdot
     E(Y\cut,\bal\cup\mu \cup\mu )\otimes E_{\bal}.\endaligned
     \tag{4.12}$$
Hence both sides of~\thetag{4.11} are isometric to ~$\Hom_{A_{\partial
Y}}\bigl( \dfrac{E_{\bal}}{\dim E_{\bal}},E(Y) \bigr)$, by \theprotag{A.7}
{Proposition}, which completes the proof.
     \enddemo

\flushpar
 Notice that in passing to the last equation in~\thetag{4.12} we trivialized
the complex line~$\Inv_{A_S}(E_\mu \otimes E_{\mu ^*})$ using the
natural duality pairing, which has norm square equal to~$\dim E_\mu $.

Next, we introduce a ring structure on~$V=\ZZ[\Phi ]$, the free abelian group
generated by the label set~$\Phi $.  A typical element of~$V$ is
denoted~$\sum\limits_{\lambda }c_\lambda \lambda $, with~$c_\lambda \in \ZZ$.
Fix $Y_3$~a 2-sphere with three open disks removed.  Give parametrizations to
the three boundary circles so that two are $-$~parametrizations and one is a
$+$~parametrization.  Set
  $$ N^\lambda _{\mu \nu } = \dim E(Y_3,\lambda \cup\mu \cup\nu ),
     \tag{4.13}$$
where $\lambda $~labels the $+$~boundary component.  Notice that
\thetag{4.13}~is independent of the choice of~$Y_3$.  Define multiplication
on~$V$ by
  $$ \mu \nu =\sum\limits_{\lambda }N^\lambda _{\mu \nu }\,\lambda .
     $$
The proof of the next proposition is quite standard.

     \proclaim{\protag{4.14} {Proposition}}
 $V$~is a commutative associative ring with identity.
     \endproclaim

\flushpar
 This ring can be used to compute $\dim E(Y,\bal)$ quite effectively.  Its
complexification~$V_{\CC}=\CC[\Phi ]$ is called the {\it Verlinde algebra\/}.

We can see the Verlinde algebra as belonging to an auxiliary
$1+1$~dimensional field theory.  Namely, for a closed oriented 1-manifold~$S$
define
  $$ \tilde{E}(S) = E(\cir\times S), \tag{4.15}$$
and for a compact oriented 2-manifold~$Y$ define
  $$ \tilde{Z}_Y = E(\cir\times Y). \tag{4.16}$$
Then it is easy to see that $\tilde{E},\tilde{Z}$ determine a
$1+1$~dimensional topological quantum field theory (as defined by the
properties in \theprotag{2.13} {Theorem}).  The Verlinde algebra is $V_{\CC}
= \tilde{E}(\cir)$.

\newpage
\head
\S{5} Computations
\endhead
\comment
lasteqno 5@ 25
\endcomment

We illustrate the theory of the previous sections with some calculations in
the quantum theory.  We begin with an arbitrary twisted theory (determined by
a cocycle~$\ah$) in $2+1$~dimensions.  Our first job is to calculate the
$SL(2;\ZZ)$~action on the vector space~$E(\cir\times \cir)$ attached to the
torus (\theprotag{5.8} {Proposition}).  One consequence is that the factor in
the gluing law~\thetag{4.11} is a matrix element~\thetag{5.12} of the
standard modular transformation~$S$.  We then take up the untwisted
($\ah=0$)~theory.  We calculate the theory explicitly in $0+1$~and
$1+1$~dimensions (\theprotag{5.17} {Proposition}).  We use the
$1+1$~dimensional theory to count the representations of a surface group into
a finite group~\thetag{5.19}.  The structure of the $2+1$~dimensional theory
on surfaces with boundary simplifies somewhat, since the central extensions
there are trivial.  The coalgebra in that theory also obtains a natural Hopf
algebra structure, which we compute directly, and the Verlinde algebra can be
derived from this Hopf algebra~(\theprotag{5.25} {Proposition}).

Fix a cocycle~$\ah\in C^3(B\Gamma ;\RZ)$.  Let $E=E(\cir\times \cir)$ be the
inner product space attached to the torus.  Then $SL(2;\ZZ)$~acts on~$E$,
by~\thetag{2.15}.  We will determine the action of the generators
  $$ T = \pmatrix 1&0\\1&1  \endpmatrix\qquad \text{and} \qquad S = \pmatrix
     0&-1\\ 1&0  \endpmatrix. \tag{5.1}$$
Cut the torus along~$\{0\}\times \cir$ to obtain the cylinder~$\zo\times
\cir$.  Note that $T$~defines a diffeomorphism of the cylinder which fixes
the boundary.  Recall that the space of fields on the cylinder is a groupoid
$\Cal{G}=\{\langle x,g\rangle :x,g\in \Gamma \}$ with composition
law~\thetag{3.10}.  The vector space of the cylinder is the space of
sections~$A$ of a line bundle $L\to\Cal{G}$, and \theprotag{3.21} {Theorem}
identifies~$E$ as the subspace of {\it central sections\/}, that is, sections
invariant under conjugation.  The coalgebra~$A$ is the direct sum
  $$ A = \bigoplus_{x,g}L_{\langle x,g  \rangle}, $$
and the dual algebra is
  $$ A^* = \bigoplus_{x,g}L^*_{\langle x,g  \rangle}, $$
as in~\thetag{4.1}.  Elements of~$A$ are then complex-valued functions
on~$A^*$.  Thus we identify~$E$ with the space of complex-valued central
functions on the algebra~$A^*$.  Since $\As$~is semisimple (\theprotag{4.2}
{Proposition}) there is a basis of character functions~$\chi _\lambda $, where
$\lambda $~runs over the set~$\Phi $ of irreducible representations.  Each
$\chi _\lambda $~is supported on
  $$ \supp \chi _\lambda =\{\langle x,g  \rangle : x\in A,\quad [x,g]=1\}
     \tag{5.2}$$
for some conjugacy class~$A\subset \Gamma $.  (Here $[x,g]=xgx\inv g\inv $ is
the group commutator.)  If we fix~$x_0\in A$, then $\chi _\lambda $~is
determined by its values on $\{\langle x_0,g \rangle : x_0g=gx_0\}$, which is
(anti)isomorphic to the centralizer~$C_x$ of~$x$ in~$\Gamma $.  The
restriction of~$\chi _\lambda $ to this set is a character
of~$\hat{C}_{x_0}$, the central extension~\thetag{3.13} of ~$C_{x_0}$.  The
Schur orthogonality relations for these characters is
  $$ \frac{1}{\#\Gamma }\int_{\Cal{G}} (\chi _\lambda ,\chi _\mu)
     = \cases 1,&\lambda =\mu ;\\0,&\text{otherwise}.\endcases \tag{5.3}$$
The integral is simply a sum over the elements of~$\Cal{G}$, since each
element of~$\Cal{G}$ has unit mass.  Here $(\cdot ,\cdot )$~is the hermitian
inner product on the line bundle~$L$.  We can rewrite the left hand side
of~\thetag{5.3} as
  $$ \sum\limits_{[x]}\sum\limits_{[g]}\frac{1}{\#(C_x\cap C_g)} \bigl( (\chi
     _\lambda )_{\langle x,g  \rangle},(\chi _\lambda )_{\langle x,g
     \rangle}\bigr), $$
where the first summation is over representatives of conjugacy classes
in~$\Gamma $ and for each~$x$ the second summation is over representatives of
conjugacy classes in~$C_x$.  But $C_g\cap C_x\cong \Aut Q$ is isomorphic to
the automorphism group of the bundle $Q\to \cir\times \cir$ obtained
from~$\langle x,g  \rangle$ by gluing.  In view of~\thetag{2.11} we have
shown the following.

     \proclaim{\protag{5.4} {Lemma}}
 The characters~$\chi _\lambda $ form an orthonormal basis of~$E$.
     \endproclaim

We have already noted~\thetag{3.12} the triviality of~$L_{\langle x,e
\rangle}$ for any~$x\in \Gamma $.  Let $\te xe\in L_{\langle x,e  \rangle}$
and $\tes xe\in L^*_{\langle x,e  \rangle}$ be the trivializing basis elements.
We also claim that there is also a trivialization
  $$ L_{\langle x,x  \rangle}\cong \CC, \tag{5.5}$$
and corresponding trivializing elements $\te xx\in L_{\langle x,x  \rangle}$
and $\tes xx\in L^*_{\langle x,x  \rangle}$.  To see this we first note
  $$ L_{\langle x,x  \rangle}\otimes L_{\langle x,x  \rangle}\cong L_{\langle
     x,x^2  \rangle} \tag{5.6}$$
by~\thetag{3.11}.    Now the diffeomorphism $T\:\zo\times \cir\to\zo\times
\cir$, defined in~\thetag{5.1}, pulls a bundle with holonomy~$\langle x,x
\rangle$ back to a bundle with holonomy~$\langle x,x^2  \rangle$, and so by
functoriality (\theprotag{3.2} {Theorem}) gives an isometry
  $$ L_{\langle x,x  \rangle}\cong L_{\langle x,x^2  \rangle}. \tag{5.7}$$
We obtain~\thetag{5.5} from~\thetag{5.6} and~\thetag{5.7}.

We claim that $\te xx$~is a central element of~$\hat{C}_x$, where
$\hat{C}_x$~is the central extension of~$C_x$ defined in~\thetag{3.13}.  For
suppose $g\in C_x$ and $\hat{\ell }\in L_{\langle x,g \rangle}$ is an element
of unit norm, so also $\hat{\ell}\in \hat{C}_x$.  Let $P_{x,x,g}\to\cir\times
\cir\times \cir$ be the $\Gamma $~bundle with holonomy~$x,x,g$ around the
three generating circles.  Then \theprotag{3.14} {Proposition} implies that
the commutator of~$\te xx$ and~$\hat{\ell}$ in~$\hat{C}_x$ is the exponential
of $\overline{F}^*[\alpha ]\bigl([\cir\times \cir\times \cir] \bigr)$, where
$F\:P_{x,x,g}\to E\Gamma $ is a classifying map.  In other words, the
commutator is the classical action evaluated on~$P_{x,x,g}$.  But
$P_{x,x,g}$~is the pullback of the bundle $\check T_{\langle x,g
\rangle}\to\cir\times \cir$ by the map
  $$ \aligned \cir\times \cir\times \cir&\longrightarrow \cir\times \cir\\
     \langle t_1,t_2,t_3 \rangle&\longmapsto \langle t_1+t_2,t_3
     \rangle,\endaligned $$
from which it follows easily that the action is~1.

Now we can compute the action of~$SL(2;\ZZ)$ on~$E$.  Let $\rho _\lambda
$~denote the representation whose character is~$\chi _\lambda $.  By the
preceding argument and Schur's lemma $\rho _\lambda (\tes xx)$~is a scalar,
if $\chi _\lambda $~is supported on~$C_x$.

     \proclaim{\protag{5.8} {Proposition}}
 The elements $T,S\in SL(2;\ZZ)$, defined in~\thetag{5.1}, act on~$E$ as
follows.  The basis~$\{\chi _\lambda \}$ diagonalizes~$T$:
  $$ T_*\chi _\lambda =\rho _\lambda (e^*_{\langle x,x \rangle})\,\chi
     _\lambda , \tag{5.9} $$
where $x$~is chosen as in~\thetag{5.2}.  Also,
  $$ \aligned
     S_*\chi _\lambda  &= \sum\limits_{\mu }\frac{1}{\#\Gamma }\int_{\Cal{G}}
     (S_*\chi _\lambda ,\chi _\mu ) \,\chi _\mu\\
     &= \frac{1}{\#\Gamma }\sum\limits_{\mu }\sum\limits_{x,g}\bigl( (\chi
     _\lambda )_{\langle g\inv,x   \rangle},(\chi _\mu )_{\langle
     x,g  \rangle}\bigr)\,  \chi _\mu .\endaligned \tag{5.10}$$
     \endproclaim

\flushpar
 Note that since $\chi _\lambda $~is a central function,
  $$ \aligned
     \chi _\lambda (\ell ^*_{\langle gxg\inv ,gxg\inv   \rangle}) &=
     \chi _\lambda \bigl( (\ell ^*_{\langle x,g  \rangle})\inv \tes xx (\ell
     ^*_{\langle x,g  \rangle}) \bigr)\\
     &= \chi _\lambda (\tes xx),\endaligned $$
where $\ell ^*_{\langle x,g  \rangle}\in L^*_{\langle x,g  \rangle}$ is any
nonzero element.  Hence \thetag{5.9}~is well-defined.
Formulas~\thetag{5.9} and~\thetag{5.10} agree with the results
in~\cite{DVVV}.

     \demo{Proof}
 The diffeomorphism~$T$ acts on the cylinder, commuting with the boundary
parametrizations, and the induced action on the set of bundles~$\Cal{G}$ is
  $$ T^*\langle x,g  \rangle = \langle x,gx  \rangle. $$
To compute the effect on the character, we can suppose $g\in C_x$ commutes
with~$x$.  Then
  $$ \langle x,gx  \rangle = \langle x,x  \rangle\cdot \langle x,g  \rangle
     $$
in the groupoid~$\Cal{G}$.  Using~\thetag{3.11} and the
trivialization~\thetag{5.5} we easily derive~\thetag{5.9}.

The gluing law in \theprotag{3.2} {Theorem} allows us to identify the induced
action of~$S$ on~$\bfldb{\cir\times \cir}$ with an action on the subset of
the groupoid~$\Cal{G}$ consisting of pairs~$\langle x,g \rangle$ of commuting
elements.  That action is
  $$ S^*\langle x,g  \rangle = \langle g\inv,x   \rangle. $$
Note that this gives an isometry
  $$ L_{\langle g\inv,x \rangle}\cong L_{\langle x,g \rangle}.  \tag{5.11} $$
Now \thetag{5.10}~follows from the Schur orthogonality
relations~\thetag{5.3}.
     \enddemo

\flushpar
 Of course, the first line of~\thetag{5.10} holds for {\it any\/} element
of~$SL(2;\ZZ)$.

There is a distinguished representation $1\in \Phi $.  Its character~$\chi
_1$ is supported on $\{\langle e,g \rangle\}\cong C_e = \Gamma $.  Now
by~\thetag{5.11} and~\thetag{3.12} we have a trivialization of $L_{\langle
e,g \rangle}\cong L_{\langle g\inv,e \rangle}$.  Then
  $$ (\chi _1)_{\langle e,g  \rangle}=e_{\langle e,g  \rangle} $$
is the corresponding trivializing element.  (This character corresponds to
the representation of $\hat{C}_e\cong \Gamma \times \TT$ which is trivial on
the $\Gamma $~factor and is standard on the $\TT$~factor.)  Now
from~\thetag{5.10} we compute the matrix element
  $$ \aligned
     (S_*\chi _\lambda ,\chi _1) &= \frac{1}{\#\Gamma }\sum\limits_{g,x}
     \bigl( (\chi _\lambda )_{\langle g\inv,x \rangle},(\chi _1)_{\langle
     x,g  \rangle}\bigr )\\
     &= \frac{1}{\#\Gamma }\sum\limits_{g}
     \bigl( (\chi _\lambda )_{\langle g\inv,e \rangle},\ell _{\langle
     e,g  \rangle}\bigr )\\
     &= \frac{1}{\#\Gamma }\sum\limits_{g\in \supp \chi _\lambda } \chi
     _\lambda (\ell ^*_{\langle g,e  \rangle})\\
     &=\frac{\dim E_\lambda }{\#\Gamma },\endaligned  \tag{5.12}$$
where $E_\lambda $~is the representation with character~$\chi _\lambda $.
This is exactly the factor which occurs in the gluing
law~\thetag{4.11}.\footnote{Walker~\cite{Wa} uses this factor (defined as
this matrix element of~$S$) in his construction of the Chern-Simons theory
with gauge group~$SU(2)$.} To avoid confusion, we point out that if $\langle
x,g \rangle\in \supp \chi _\lambda $, and there are $k$~elements in the
conjugacy class of~$x$, then $\dim E_\lambda =k\cdot \dim
\overline{E}_\lambda $, where $\overline{E}_\lambda $~is the irreducible
representation of~$\hat{C}_x$ with character~$\chi _\lambda \res{\hat{C}_x}$.

We turn now to the untwisted ($\ah=0$) theory.  The quantum theory described
in~\S{2} makes sense in arbitrary dimensions (and the manifolds do not have
to be oriented).  Formulas \thetag{2.6}--\thetag{2.8} can be worked out
more explicitly.  First, for any closed, connected $d$-manifold~$Y$ we
use~\thetag{2.7} and~\thetag{1.1} to identify
  $$ E(Y) = L^2(\Hom(\pi _1Y,\Gamma )/\Gamma ), $$
where the mass of a representation $\gamma \in \Hom(\pi _1Y,\Gamma )$ is
  $$ \mu _{[\gamma ]} = \frac{1}{\#C_\gamma }. \tag{5.13}$$
Here $C_\gamma $~is the centralizer of the image of~$\gamma $ in~$\Gamma $
and $[\gamma ]$~denotes the set of conjugates of a representation~$\gamma $.
For a closed, connected $(d+1)$-manifold~$X$ we compute from~\thetag{2.6}:
  $$ \spreadlines{6pt}
     \aligned
     Z_X &= \sum\limits_{[\beta ]\in \Hom(\pi _1X,\Gamma )/\Gamma }
     \frac{1}{\#C_\beta } \\
     &= \sum\limits_{\beta \in \Hom(\pi _1X,\Gamma )}
     \frac{1}{\#C_\beta } \,\#[\beta ]\\
     &= \frac{1}{\#\Gamma }\,\#\Hom(\pi _1X,\Gamma ).\endaligned \tag{5.14}$$
If $X$~is a compact, connected $(d+1)$-manifold with nonempty connected
boundary, and $\gamma \in \HG {\partial X}$, then from~\thetag{2.8} we
deduce
  $$ Z_X(\gamma ) = \#(\iota ^*)\inv (\gamma ), $$
where $\iota ^*\:\HG X\to\HG{\partial X}$ is the restriction map.

The theory in $0+1$~dimensions is completely trivial.  The inner product
space $E(\ptt)\cong \frac{1}{\#\Gamma }\cdot \CC$ and
$Z_{[0,1]}\:\frac{1}{\#\Gamma }\cdot \CC\to \frac{1}{\#\Gamma }\cdot \CC$ is
the identity map.  From~\thetag{5.14} we have $Z_{S^1}=1$.  Notice that this
is consistent with the gluing law~\thetag{2.17} applied to~$[0,1]$.

In $1+1$~dimensions the inner product space $E=E(\cir)$ of the circle carries
extra structure.  Let $Y_3$~denote $S^2$ with three disks removed.
Parametrize the boundary circles so that one is positively oriented and the
remaining two are negatively oriented.  Notice that
  $$ Z_{Y_3}\:E\otimes E\longrightarrow E \tag{5.15}$$
is independent of the parametrizations, since the space of orientation
preserving diffeomorphisms of~$\cir$ is connected.  (We use the inner product
on~$E$ to write~$Z_{Y_3}$ in this form---cf\.  \theprotag{2.13(b)}
{Theorem}.)  It is easy to verify from the axioms that \thetag{5.15}~defines
an algebra structure on~$E$ which is commutative and associative.
Furthermore, if $Y_1$~denotes a disk with positively parametrized boundary,
then $Z_{Y_1}\in E$ is an identity element for this multiplication.  The
algebra~$E$ has an involution $a\mapsto \bar a$ induced by any orientation
reversing diffeomorphism of~$\cir$, and it is unitary in the sense dual
to~\thetag{A.3}:
  $$ (a,bc) = (a\bar{c},b) = (\bar{b}a,c) $$
for all $a,b,c\in E$.  It now follows that $E$~is a semisimple algebra, and
so is isomorphic to a direct product $\CC\times \cdots\times \CC$.  Up to
isomorphism this unitary algebra is determined by the norm squares $\lambda
_i^2>0$ of the identity element in each factor of the direct product.

The preceding discussion applies to any $1+1$~dimensional unitary topological
field theory (in the sense of \theprotag{2.13} {Theorem}).  Conversely, such
theories can be constructed from unitary commutative, associative algebras
with unit.  We compute the partition function~$Z_{Y(g)}$ of a closed,
oriented surface of genus~$g$ in terms of this algebra:
  $$ Z_{Y(g)} = \sum\limits_{}\bigl(\lambda _i^2\bigr)^{1-g}. \tag{5.16}$$
These considerations apply to the Verlinde algebra
(cf\.~\thetag{4.15}--\thetag{4.16}); in that context \thetag{5.16}~appears
in~\cite{V}.

Returning to the untwisted theory attached to a finite group we have the
following.

     \proclaim{\protag{5.17} {Proposition}}
 In the untwisted $1+1$-dimensional theory $E=E(\cir)$ is naturally identified
with the character ring~$\Cal{F}_{\!\!\text{cent}}(\Gamma)$ of complex-valued
central functions on~$\Gamma $.  The multiplication is by convolution.  The
hermitian structure is
  $$ (f _1,f _2) = \frac{1}{\#\Gamma }\sum\limits_{x\in \Gamma }f
     _1(x)\overline{f _2(x)}, \qquad \chi _1,\chi _2\in
     \Cal{F}_{\!\!\text{cent}}(\Gamma).  \tag{5.18} $$
Finally,
  $$ Z_{Y(g)} = \frac{1}{\#\Gamma }\#\Hom( \pi _1Y(g),\Gamma ) = (\#\Gamma
     )^{2g-2}\sum\limits_{i}(\dim E_i)^{2-2g}, \tag{5.19}$$
where $i$~runs over the irreducible representations~$E_i$ of~$\Gamma $.
     \endproclaim

\flushpar
 Formula~\thetag{5.19}, which counts the number of representations of a
surface group in a finite group, can also be derived using standard methods
in finite group theory~\cite{Se}.  Here we derive it by chopping~$Y(g)$ into
a union of ``pairs of pants'' and annuli.  This is a simple illustration of
how gluing laws are used in topological quantum field theory to compute
global invariants from local computations.


     \demo{Proof}
 The equivalence classes ~$\fldb{\cir}$ of bundles over the circle correspond
to conjugacy classes in~$\Gamma $, and for~$g\in \Gamma $ the conjugacy class
of~$g$ is weighted by~$1/\#C_g$, according to~\thetag{5.13}.  Now
\thetag{5.18}~follows immediately.  To compute the
multiplication~\thetag{5.15} consider $Y_3$ with a basepoint on each boundary
component.  Then, as in Figure~1, the set of equivalence
classes~$\bfldb{Y_3}$ is in $1:1$~correspondence with the set of 4-tuples
$x_1,g_1,x_2,g_2\in \Gamma $.  In other words, there is a 1:1~correspondence
  $$ \bfldb{Y_3} \longleftrightarrow \Cal{G}\times \Cal{G}.\tag{5.20}$$
The path integral over~$Y_3$ is defined in~\thetag{2.12}.  Then that fact
that $E$~consists of central functions implies that the
multiplication~\thetag{5.15} is the convolution
  $$ (f _1*f _2)(x) = \sum\limits_{x_1x_2=x} f _1(x_1)f _2(x_2).
     $$
Note that the factor of~$(\#\Gamma )^2$ which comes from the summation
over~$g_1,g_2$ is canceled by the factor in the inner product~\thetag{5.18}.
The characters of the irreducible representations, suitably normalized, are
commuting idempotents in~$\Cal{F}_{\!\!\text{cent}}(\Gamma)$, by the Schur
orthogonality relations.  Hence
  $$ \lambda ^2_i = \(\frac{\dim E_i}{\#\Gamma }\)^2, $$
from which \thetag{5.19}~follows immediately given~\thetag{5.16}.
     \enddemo

The modular structure of~\S{3} and ~\S{4} has an analog here: Consider
bundles over the interval with basepoints over the endpoints.  The reader
should check the assertions in \theprotag{3.21} {Theorem} and \theprotag{4.9}
{Proposition} in this $1+1$~dimensional theory.  They all reduce to standard
facts about finite groups.  The basic coalgebra~$A$ is simply the coalgebra
of functions on~$\Gamma $.  The twisted theory in $1+1$~dimensions can also
be made quite explicit.  Namely, a cocycle $\ah \in C^2(B\Gamma ;\RR/\ZZ)$
determines a central extension
  $$ 1 @>>> \TT @>>> \hat{\Gamma} @>>> \Gamma @>>> 1. $$
Then the basic algebra~$A_{\ah}$ in the corresponding twisted quantum theory
is the coalgebra of functions twisted by the cocycle~$\ah$.  The
representations of this coalgebra correspond to representations
of~$\hat{\Gamma }$ which are standard on the central~$\TT$.  The reader may
wish to work out this twisted case in detail.

Finally, we take up the modular structure in the {\it untwisted\/}
$2+1$~dimensional theory.  Consider once more~$Y_3$, as shown in Figure~1,
with two boundary circles negatively oriented, one boundary circle positively
oriented, and basepoints on each of the boundary circles.
Quantizing~\thetag{5.20} we find a vector space isomorphism
  $$ E(Y_3) \cong A\otimes A, \tag{5.21}$$
where $A=E(\zo\times \cir)$ is the unitary coalgebra of the untwisted theory.
Recall that this is just the set of functions on~$\Cal{G}$.  Now $E(Y_3)$~has
a right $A$-comodule structure~$\Delta $ coming from the positively oriented
boundary circle.  Combining with the counit~\thetag{3.29} we obtain
a map
  $$ m\:A\otimes A @>\Delta >> A\otimes A\otimes A @> \epsilon \otimes
     \epsilon >> A. $$
More explicitly, if $a,b\in A$ are functions on~$\Cal{G}$, then
  $$ m(a,b)_{\langle x,g  \rangle} = \sum\limits_{x_1x_2=x}a_{\langle x_1,g
     \rangle}b_{\langle x_2,g\rangle}. $$
Also, there is a canonical element $\bone\in A$ as follows.  First, if
$Y_1$~is the disk then $E(Y_1)\cong \CC$ is the trivial comodule.  But we can
write~$Y_1$ as the union of~$Y_1$ and $Y_2=\zo\times \cir$, and
by~\thetag{3.26} this gives an isometry
  $$ \CC \boxtimes A\cong \CC. $$
{}From this we determine~$\bone\in A$:
  $$ \bone_{\langle x,g  \rangle} = \cases 1
     ,&x=e;\\0,&\text{otherwise}.\endcases $$

     \proclaim{\protag{5.22} {Proposition}}
 The multiplication~$m$ and unit~$\bone$ render the coalgebra~$A$ a Hopf
algebra.
     \endproclaim

\flushpar
 This Hopf algebra is the dual of the ``quantum double'' of~$\Gamma $
considered in~\cite{DPR}.  We leave the verification of the Hopf algebra
axioms to the reader.

The algebra structure can be used to define tensor products of
comodules of~$A$.  Namely, if $E_1,E_2$ are right comodules, with
coproducts~$\Delta _1,\Delta _2$, then the composition
  $$ E_1\otimes E_2 @>\Delta _1\otimes \Delta _2>> E_1\otimes A\otimes
     E_2\otimes A @>>> E_1\otimes E_2\otimes A\otimes A @>1\otimes 1\otimes
     m>> E_1\otimes E_2\otimes A \tag{5.23}$$
is a coproduct on~$E_1\otimes E_2$.  We denote this comodule by~$E_1\otimes
_AE_2$.  In particular, we can take tensor products of the irreducible
corepresentations.  Define nonnegative integers~$\tilde{N}^\lambda _{\mu \nu
}$ by
  $$ E_{\mu }\otimes _AE_\nu \cong \bigoplus_\lambda \tilde{N}^\lambda _{\mu
     \nu }\,E_\lambda . \tag{5.24}$$

     \proclaim{\protag{5.25} {Proposition}}
 We have
  $$ \tilde{N}^\lambda _{\mu \nu } = N^\lambda _{\mu \nu }, $$
where $N^\lambda _{\mu \nu }$ ~is defined in~\thetag{4.13}.
     \endproclaim

\flushpar
 In other words, this tensor product on corepresentations reproduces the
Verlinde algebra.

     \demo{Proof}
 Recall that $E(Y_3)$~is a right $A\op\times A\op\times A$-comodule.  Further,
the isomorphism~\thetag{5.21} is defined in terms of the
identification~\thetag{5.20}.  Under this isomorphism $A\otimes A$ becomes a
right $A\op\times A\op\times A$-comodule.  The right $A\op\times
A\op$-comodule structure is simply the natural {\it left\/} $A\times
A$-comodule structure on~$A\otimes A$; the right $A$-comodule structure is
that of the tensor product~$A\otimes _AA$, as defined by~\thetag{5.23}.
{}From~\thetag{4.7}, \thetag{4.13}, and~\thetag{A.9} we compute
  $$ \split
     N^\lambda _{\mu \nu } &= \dim E(Y_3,\lambda \cup \mu \cup \nu )\\
     &= \dim \Hom_{A\op\times A\op\times A}(E_\mu ^*\otimes E_{\nu }^*\otimes
     E\mstrut _\lambda , A\otimes A)\\
     &= \dim \Hom_A(E_\lambda ,E_\mu \otimes _AE_\nu ).\endsplit $$
But this last expression is exactly~$\tilde{N}^\lambda _{\mu \nu }$,
from the definition~\thetag{5.24}.
     \enddemo

A much more complete treatment of the Hopf algebra structure in both the
untwisted and twisted theories appears in~\cite{F5}.

\newpage
\head
\S{A} Appendix: Coalgebras and Comodules
\endhead
\comment
lasteqno A@  9
\endcomment

Some basic definitions may be found in~\cite{MM,\S2}.  A {\it
coalgebra\/}~$A$ over~$\CC$ is a complex vector space endowed with a {\it
comultiplication\/}
  $$ \Delta \:A\longrightarrow A\otimes A  \tag{A.1}$$
and a {\it counit\/}
  $$ \epsilon \:A\longrightarrow \CC. \tag{A.2}$$
The comultiplication is required to be coassociative.  A (right) {\it
comodule\/}~$E$ is a complex vector space with a coaction
  $$ \Delta \:E\longrightarrow E\otimes A $$
which is compatible with~\thetag{A.1} and~\thetag{A.2}.

A natural example is the following.  Suppose $G $~is a finite group acting on
a finite set~$X$.  Then the vector space~$\Cal{F}(G )$ of complex-valued
functions on~$G $ is a coalgebra and the space of functions~$\Cal{F}(X)$ is a
comodule.  The comultiplication is dual to the group multiplication $G \times
G \to G $ and the coaction is dual to the group action $X\times G \to X$.
The counit is $\epsilon (f)=f(1)$, where $1\in G $~is the identity element.

Now suppose $a\mapsto \bar{a}$ is an antiinvolution of~$A$.  Then $A$~ is
{\it unitary\/} if $A$~is endowed with a hermitian inner product which
satisfies
  $$ (\Delta a,b\otimes c) = (a \otimes \bar{c},\Delta b) = (\bar{b}\otimes
     a,\Delta c) \tag{A.3}$$
for all $a,b,c\in A$.  A right comodule~$E$ is {\it unitary\/} if it has a
hermitian inner product which satisfies
  $$ (\Delta x,y\otimes a) = (x \otimes \bar{a},\Delta y) \tag{A.4}$$
for all $a\in A$, $x,y\in E$.

Suppose $E_R$~is a right $A$-comodule and $E_L$~is a left $A$-comodule.  Then
the {\it cotensor product\/}~$E_R \boxtimes_A E_L$ is the vector subspace
of~$E_R \otimes _{\CC}E_L$ annihilated by~$\Delta_R \otimes \id - \id \otimes
\Delta_L $.  It is not a comodule, but rather is simply a vector space.  If
$E_R,E_L$~are unitary, then $E_R \boxtimes_A E_L$ inherits the subspace inner
product.  More generally, if $E$~has both a left $A$-comodule
structure~$\Delta _L$ and a right $A$-comodule structure~$\Delta _R$, then we
define
  $$ \Inv_A(E)\subset E $$
to be the subspace annihilated by~$\Delta _L - P\Delta _R$, where
$P\:E\otimes _\CC A \to A \otimes _{\CC}E$ is the natural isomorphism.

The dual~$A^*$ of a coalgebra is an algebra.  In finite dimensions the dual
of an algebra is a coalgebra, so we can pass freely between the two.  Recall
that an algebra (over~$\CC$) is {\it simple\/} if it is isomorphic to a
matrix algebra, and it is {\it semisimple\/} if it is isomorphic to a direct
product of matrix algebras.  If
  $$ \As\cong \prod\limits_{\lambda \in \Phi }M_\lambda   \tag{A.5}$$
is a finite direct product of matrix algebras~$M_\lambda $, then there are
commuting idempotents $\as_\lambda \in \As$ which correspond to the identity
matrix in~$M_\lambda $.  Now each $M_\lambda $ has a unique nonzero irreducible
representation~$\Es_\lambda $ (up to isomorphism).  Under the
isomorphism~\thetag{A.5} we view~$\Eas$ as an irreducible representation
of~$\As$.  Then if $\Es$~is any representation of~$\As$, there is an
isomorphism
  $$ \aligned
     \bigoplus_{\lambda \in \Phi } \Hom_{\As}(\Eas,\Es)\otimes \Eas
     &\longrightarrow \Es\\
     \oplus_\lambda  (f\mstrut _\lambda \otimes \es_\lambda) &\longmapsto
     \sum\limits_{\lambda }f\mstrut _\lambda (\es_\lambda ).\endaligned
     \tag{A.6}$$

Now suppose that $\As$~is a unitary algebra, i.e., has a hermitian inner
product which satisfies the dual of~\thetag{A.3}:
  $$ (\as,\bs\cs) = (\as\overline{\cs},\bs) = (\overline{\bs}\as,\cs).
     $$
Then an easy argument with the ~$\as_\lambda $ shows that the images
of~$M_\lambda $ in~$\As$ under the isomorphism~\thetag{A.5} are orthogonal.
Since each matrix algebra~$M_\lambda $ has a unique unitary structure up to
scaling, we may assume that \thetag{A.5}~is an isometry.  The irreducible
module~$\Eas$ also has a unique unitary structure up to scaling.  Fix unitary
structures on the~$\Eas$.  Now suppose that $\Es$~is a unitary $\As$-module,
and $f,f'\in \Hom_{\As}(\Eas,\Es)$.  Then for $e^*,\esp\in \Eas$ we set
  $$ h(\es,\esp) = \bigl(f(\es),f'(\esp)\bigr)_{\Es}. $$
This defines a new unitary structure on~$\Eas$.  So by uniqueness,
  $$ h(\es,\esp) = \lambda \cdot (\es,\esp)_{\Eas} $$
for some constant~$\lambda $.  Let $\{\es_i\}$~be an orthonormal basis
of~$\Eas$.  Then the inner product~$(f,f')$ in~$\Hom_{\As}(\Eas,\Es)$ is
  $$ \aligned
     (f,f') &= \sum\limits_{\es_i}\bigl( f(\es_i),f'(\es_i)\bigr)_{\Es}\\
     &= \lambda \cdot \dim \Eas.\endaligned $$
Using this we see that \thetag{A.6}~is an isometry if we scale the inner
product on~$\Eas$ in~$\Hom_{\As}(\Eas,\Es)$ by a factor~$1/\dim\Eas$.
Dually, we have proved the following.

     \proclaim{\protag{A.7} {Proposition}}
 Suppose $A$~is a finite dimensional unitary semisimple coalgebra, and
$\{E_\lambda \}_{\lambda \in \Phi }$ a representative set of irreducible
corepresentations.  Fix unitary structures on each ~$E_{\lambda }$.  Then for
any unitary comodule~$E$ the map
  $$ \aligned
     \bigoplus_{\lambda \in \Phi }\Hom_{A}(\frac{E_\lambda }{\dim E_\lambda
     },E)\otimes E_\lambda  &\longrightarrow E\\
     \oplus_\lambda  (f_\lambda \otimes e_\lambda) &\longmapsto
     \sum\limits_{\lambda }f_\lambda (e_\lambda ).\endaligned  \tag{A.8}$$
is an isometry.
     \endproclaim

\flushpar
 (Recall that $\dfrac{E_\lambda }{\dim E_\lambda }$ denotes the
space~$E_\lambda $ with the inner product scaled by the factor~$1/\dim
E_\lambda $.)

If $E$~is a right $A$-comodule, then $\Es$~has a left $A$-comodule structure
determined by the formula
  $$ \langle e,\Delta _{\Es}\es\rangle =\langle \Delta _Ee,\es\rangle \in
     A,\qquad e\in E,\quad \es \in\Es. $$
If $E$~is unitary, then so is~$\Es$.  In this way $\{\Eas\}$~is a
representative list of the irreducible left $A$-comodules.  Equivalently, it
is a representative list of the irreducible right comodules for the opposite
coalgebra~$A\op$.

{}From~\thetag{A.5} it follows that a semisimple coalgebra~$A$ decomposes as
  $$ A\cong \bigoplus_{\lambda \in \Phi }E_\lambda ^*\otimes E\mstrut
     _\lambda . \tag{A.9}$$
We can interpret~\thetag{A.9} as an equation for right, left, or bi
$A$-comodules.

Finally, for $\lambda ,\mu \in \Phi $ we have the formula
  $$ \Inv_A(E\mstrut _\lambda \otimes \Es_\mu ) = E\mstrut _\lambda
     \boxtimes_A \Es_\mu
     \cong \cases \dim E_\lambda \cdot \CC ,&\text{if $\lambda =\mu $}
     ;\\0,&\text{if $\lambda \not= \mu $} .\endcases  $$
The factor $\dim E_\lambda $ is the norm square of the canonical element (the
duality pairing) in~$E\mstrut _\lambda \otimes \Es_\lambda $.

\newpage
\head
\S{B} Appendix: Integration of Singular Cocycles
\endhead
\comment
lasteqno B@ 11
\endcomment

Fix an integer~$d$, and suppose that $X$ is a compact oriented
$(d+1)$-manifold with boundary.  Than if $\alpha $~is a differential form of
degree~$d+1$ on~$X$, the integral~$\int_{X}\alpha $ is well-defined.  Notice
that~$d\alpha =0$ since any $(d+2)$-form on~$X$ vanishes.  If instead we
consider a (real-valued) singular cocycle~$\alpha $, then the
integral~$\int_{X}\alpha $~is well-defined if $X$~is closed.  For in this
case $X$~has a fundamental class $[X]\in H_{d+1}(X)$ and the integral is the
pairing of~$[X]$ with the cohomology class represented by~$\alpha $.  If
$\partial X\not= 0$ we must work a little harder, essentially because $\alpha
$~may not vanish on degenerate chains.  (By contrast differential forms
vanish on degenerate chains.)  Our constructions in this appendix keep track
of these degeneracies.  Notice that we only define integration of {\it
closed\/} cochains, i.e., cocycles.  We work first with manifolds of
arbitrary dimensions~$d$ and~$d+1$, though there is a generalization to
CW~complexes.  Then we specialize to~$d=2$ and extend the theory to surfaces
with boundary by fixing some standard choices on the boundary.

     \proclaim{\protag{B.1} {Proposition}}
 Let $Y$~be a closed oriented $d$-manifold and $\alpha \in C^{d+1}(Y;\RZ)$ a
singular cocycle.  Then there is a metrized ``integration line''~$I_{Y,\alpha
}$ defined.  If $X$~is a compact oriented $(d+1)$-manifold, $i\:\partial
X\hookrightarrow X$ the inclusion of the boundary, and $\alpha \in
C^{d+1}(X;\RZ)$ a cocycle, then
  $$ \eint X{\alpha }\in I_{\partial X,i^*\alpha } $$
is defined and has unit norm.  These lines and integrals satisfy:\newline
 \rom(a\rom)\ \rom({\it Functoriality\/}\rom)\ If $f\:Y'\to Y$ is an
orientation preserving diffeomorphism, then there is an induced isometry
  $$ f_*\:\intline {Y'}{f^*\alpha }  \longrightarrow \iline{Y} $$
and these compose properly.  If $F \:X'\to X$ is an orientation preserving
diffeomorphism, then
  $$ (\partial F)_*\left[\eint {X'}{F^*\alpha } \right] = \eint {X}{\alpha
     }. $$
 \rom(b\rom)\ \rom({\it Orientation\/}\rom)\ There is a natural isometry
  $$ \iline{-Y} \cong \overline{\iline{Y}},  $$
and
  $$ \eint{-X}{\alpha }   = \overline{\eint X{\alpha }}. $$
 \rom(c\rom)\ \rom({\it Additivity\/}\rom)\ If $Y=Y_1\sqcup Y_2$ is a disjoint
union, then there is a natural isometry
  $$ \intline{Y_1\sqcup Y_2}{ \alpha _1\sqcup  \alpha _2} \cong
     \intline{Y_1}{ \alpha _1}\otimes \intline{Y_2}{ \alpha _2}. $$
If $X=X_1\sqcup X_2$ is a disjoint union, then
  $$ \eint{X_1\sqcup X_2}{\alpha _1\sqcup \alpha _2} =
     \eint{X_1}{\alpha _1} \otimes \eint{X_2}{\alpha _2}.   $$
 \rom(d\rom)\ \rom({\it Gluing\/}\rom)\ Suppose $j\:Y\hookrightarrow X$ is a
closed oriented codimension one submanifold and $X\cut$~is the manifold
obtained by cutting~$X$ along~$Y$.  Then $\partial X\cut = \partial X\sqcup Y
\sqcup -Y$.  Suppose $\alpha \in C^{d+1}(X;\RZ)$~is a singular
$(d+1)$-cocycle on~$S$, and $\alpha \cut \in C^{d+1}(X\cut;\RZ)$~the induced
cocycle on~$X\cut$.  Then
  $$ \eint X{\alpha } = \Tr_{Y,j^* \alpha
     }\left[\eint{X\cut}{\alpha \cut}\right],
     \tag{B.2}$$
where $\Tr_{Y,j^* \alpha  }$ is the contraction
  $$ \Tr_{Y,j^* \alpha } \:\intline{\partial X\cut}{ \alpha \cut}
     \cong \intline{\partial X}{i^* \alpha  }\otimes \intline{Y}{j^* \alpha }
     \otimes \overline{\intline{Y}{j^* \alpha }}\longrightarrow
      \intline{\partial X}{i^* \alpha  } $$
using the hermitian metric on~$\intline{Y}{j^* \alpha } $.\newline
 \rom(e\rom)\ \rom({\it Stokes' Theorem \rom I\/}\rom)\ Let $\alpha \in
C^{d+1}(W;\RZ)$~be a singular cocycle on a compact oriented
$(d+2)$-manifold~$W$.  Then
  $$ \eint{\partial W}{\alpha } = 1. \tag{B.3}$$
 \rom(f\rom)\ \rom({\it Stokes' Theorem \rom{II}\/}\rom)\ A singular
$d$-cochain~$\beta \in C^{d}(Y;\RZ)$ on~$Y$ determines a trivialization
  $$ \intline{Y}{\delta \beta }\cong \CC.   $$
A singular $d$-cochain~$\beta \in C^{d}(X;\RZ)$ on~$X$ satisfies
  $$ \exp\(\tpi \int_{X}\delta \beta \) = 1  $$
under this isomorphism.
     \endproclaim

     \demo{Proof}
 We give the constructions and leave the reader to verify the properties.
Consider the category~$\Cal{C}_Y$ whose objects are oriented cycles $y\in
C_{d}(Y)$ which represent the fundamental class $[Y]\in H_{d}(Y)$.  A
morphism $y @>a>> y'$ is a chain~$a\in C_{d+1}(Y)$ such that $y'=y+\partial a$.
Define a functor $\Cal{F}_{Y,\alpha }\:\Cal{C}_Y\to\Cal{L}$
by~$\Cal{F}_{Y,\alpha }(y)=\CC$ for each object ~$y$ and $\Cal{F}_{Y,\alpha
}(y@>a>> y')$ acts as multiplication by $e^{2\pi i\alpha (a)}$.  Now if $y
@>a>> y$ is an automorphism, then ~$\partial a=0$.  Since $H_{d+1}(Y)=0$, there
is a $(d+2)$-chain~$b$ with~$\partial b=a$.  Hence $\alpha (a) = \alpha
(\partial b) = \delta \alpha (b)=0$.  Therefore, the functor~$\Cal{F}_Y$ has
no holonomy, so defines the desired line~$I_{Y,\alpha }$ of invariant
sections.

For~$X$, choose a chain $x\in C_{d+1}(X)$ which represents the fundamental
class $[X]\in H_{d+1}(X,\partial X)$.  Then $\partial x\in C_{d}(\partial X)$
is closed and represents the fundamental class $[\partial X]\in
H_{d}(\partial X)$.  Consider the section
  $$ \partial x\longmapsto e^{2\pi i\alpha (x)} \tag{B.4}$$
of the functor~$\Cal{F}_{\partial X,i^*\alpha }$.  If $x'$~is another chain
representing~$[X]$ with~$\partial x=\partial x'$, then $x'=x + \partial c$
for some $c\in C_{d+1}(X)$.  But then $\alpha (\partial c) = \delta \alpha (c)
= 0$, so that \thetag{B.4}~is well-defined.  A similar check shows that
\thetag{B.4}~is an invariant section of~$\Cal{F}_{\partial X,i^*\alpha
}$, so determines an element of unit norm in~$I_{\partial X,i^*\alpha }$ as
desired.
     \enddemo

We generalize these constructions\footnote{There is a simpler version of what
follows for~$d=1$.  In that case there is no need to fix a standard cycle~$s$
nor to parametrize the boundary.  The analogue of \theprotag{B.5}
{Proposition} holds, now for gluings of intervals.  We use the $d=1$~version
in the proof of \theprotag{3.14} {Proposition}.} in the case~$d=2$.  Fix once
and for all the standard oriented circle~$\cir = \zo\bigm/0\sim 1$ and the
standard cycle $s\in C_1(\cir)$ which represents the fundamental
class~$[\cir]$. (Thus $s$~is the identity map $\zo\to\zo$ followed by the
quotient map onto~$\cir$.)  The following proposition generalizes the
construction of integration lines to surfaces with boundary.

     \proclaim{\protag{B.5} {Proposition}}
 Let $Y$~be a compact oriented 2-manifold, and suppose that each
component~$(\partial Y)_i$ of the boundary is endowed with a fixed
parametrization $\cir\to(\partial Y)_i$ (which may or may not preserve the
orientation).  Suppose $\alpha \in C^3(Y;\RZ)$ is a singular cocycle.  Then
there is a metrized line~$I_{Y,\alpha }$ defined.  If~$\partial Y=\emptyset
$, then this is the line defined in \theprotag{B.1} {Proposition}.  These
lines satisfy properties \rom(a\rom)\footnote{The diffeomorphisms should
commute with the boundary parametrizations.}, \rom(b\rom), \rom(c\rom) of
that proposition and in addition satisfy:\newline
 \rom(d\rom)\ \rom({\it Gluing\/}\rom)\ Suppose $S\hookrightarrow Y$ is a
closed embedded 1-manifold and $Y\cut$~the manifold obtained by cutting
along~$S$.  Then $\partial Y\cut = \partial Y \sqcup S\sqcup -S$ and we use
parametrizations which agree on~$S$ and~$-S$.  Let $\alpha $~be a $3$-cocycle
on~$Y$ and $\alpha \cut$~the induced cocycle on~$Y\cut$.  Then there is an
isometry
  $$ I_{Y,\alpha } \cong I_{Y\cut,\alpha \cut}. \tag{B.6}$$
These isometries compose properly under successive gluings.
     \endproclaim

     \demo{Proof}
 The construction is similar to that in \theprotag{B.1} {Proposition}.  Using
the boundary parametrizations we construct from~$s$ and~$-s$ a cycle $z\in
C_1(\partial Y)$ which represents the fundamental class~$[\partial Y]$.  We
take $\Cal{C}_Y$~to be the category whose objects are oriented cycles $y\in
C_2(Y)$ which represent the fundamental class~$[Y,\partial Y]$ and
satisfy~$\partial y=z$.  A morphism $y @>a>> y'$ is a chain $a\in C_3(Y)$
with $y' = y + \partial a$.  Notice that $\Cal{C}_Y$~is connected.  The rest
of the construction of~$I_{Y,\alpha }$ is as before.

For the gluing law we proceed as follows.  The gluing map $g\:Y\cut\to Y$
induces a map $g_*\:C_2(Y\cut)\to C_2(Y)$ on chains, and this in turn induces
a map $g_*\:\Cal{C}_{Y\cut}\to \Cal{C}_Y$ since the boundaries of the chains
along~$S$ and~$-S$ cancel out under gluing.  Then $g_*$~extends in an obvious
way to a functor, and $\Cal{F}_{Y\cut,\alpha \cut} = \Cal{F}_{Y,\alpha }\circ
g_*$.  Thus $g_*$~induces the desired isometry on the space of invariant
sections.
     \enddemo

Suppose $\beta \in C^3(\cir;\RZ)$ is a singular cocycle.  We also use~ `$\beta
$' to denote the induced cocycles on $\zo\times \cir$ and $\cir\times \cir$,
obtained by pullback from the second factor.  Glue two cylinders together to
form a single cylinder.  Then \thetag{B.6}~leads to an isomorphism
  $$ I_{\zo\times \cir,\beta }\otimes I_{\zo\times \cir,\beta }\cong
     I_{\zo\times \cir,\beta }. \tag{B.7}$$
There is a unique trivialization
  $$ I_{\zo\times \cir,\beta }\cong \CC \tag{B.8}$$
which is compatible with~\thetag{B.7}.  Gluing the two ends of $\zo\times
\cir$ together and applying~\thetag{B.6} we obtain a trivialization
  $$ I_{\cir\times \cir,\beta }\cong \CC \tag{B.9}$$
compatible with~\thetag{B.8} under gluing.

Finally, we observe that the gluing law in \theprotag{B.1} {Proposition}(d)
extends to a 3-manifold glued along part of its boundary.

     \proclaim{\protag{B.10} {Proposition}}
 Let $X$~be a compact oriented 3-manifold and $Y$~ a compact oriented
2-manifold with parametrized boundary.  Suppose $Y\hookrightarrow X$ is an
embedding which restricts to an embedding $\partial Y\hookrightarrow \partial
X$. Let $X\cut$~be the ``manifold with corners'' obtained by cutting~$X$
along~$Y$.  Then $\partial X\cut = \partial X\cup Y \cup -Y$ where the union
is over~$\partial Y\sqcup -\partial Y$.  Suppose $\alpha \in C^{d+1}(X;\RZ)$~is
a $3$-cocycle, with $\alpha \cut$~the induced cocycle on~$X\cut$, and $\alpha
$~the restriction of~$\alpha $ to~$Y$.  Then
  $$ \exp\(\tpi\int_X\alpha\)  = \Tr_{Y,\alpha }\left[\exp \(
     \tpi\int_{X\cut}{\alpha \cut}\)\right],  $$
where $\Tr_{Y,\alpha }$ is the contraction
  $$ \Tr_{Y,\alpha } \:I_{\partial X\cut, \alpha \cut} \cong  I_{\partial X,
     \alpha }\otimes I_{Y,\alpha }  \otimes \overline{I_{Y,\alpha
     }}\longrightarrow I_{\partial X,\alpha }  \tag{B.11}$$
using the hermitian metric on~$I_{Y,\alpha } $.
     \endproclaim

\flushpar
 There is a canonical way to ``straighten the angle''~\cite{CF} to
make~$X\cut$ a smooth manifold with boundary.  In particular, $X\cut$~has a
(relative) fundamental class, which is needed to define the integration lines
and the integrals. Note that \thetag{B.11}~uses the
isomorphism~\thetag{B.6}.  We have also implicitly used~\thetag{B.6} when
we cut~$\partial X$ along~$\partial Y$.  The proof of \theprotag{B.10}
{Proposition} is straightforward once the definitions are clear.

\vfill

\centerline{\bf Figure Caption}
\bigskip
\noindent{Figure~1: The bundle over~$Y_3$ corresponding to $\langle x_1,g_1
\rangle\times \langle x_2,g_2  \rangle\in \Cal{G}\times\Cal{G}$}

\enddocument